\documentclass[aps, superscriptaddress, preprint]{revtex4}
\usepackage{amsmath}
\usepackage{amsfonts}
\usepackage{amssymb}
\usepackage{graphicx}
\usepackage{units}
\usepackage{csquotes}
\usepackage{color}
\usepackage{xspace}
\usepackage[normalem]{ulem}
\usepackage{natbib}

\newcommand{\fig}[1]{Fig.~\ref{#1}}

\begin{document}

\title{Anisotropic Magnetodielectric Coupling in Layered Antiferromagnetic FePS$_3$}

\author{Anudeepa Ghosh}
\affiliation{School of Physical Sciences, Indian Association for the Cultivation of Science, 2A \& B
Raja S. C. Mullick Road, Jadavpur, Kolkata - 700032, India
}
\author{Magdalena Birowska}
\affiliation{Faculty of Physics, Institute of Theoretical Physics, University of Warsaw, Pasteura 5, 02093, Warsaw, Poland}

\author{Pradeepta Kumar Ghose}
\affiliation{School of Physical Sciences, Indian Association for the Cultivation of Science, 2A \& B
Raja S. C. Mullick Road, Jadavpur, Kolkata - 700032, India
}

\author{Miłosz Rybak}
\affiliation{Department of Semiconductor Materials Engineering, Faculty of Fundamental Problems of Technology, Wrocław University of Science and Technology, Wybrzeże Wyspiańskiego 27, PL-50370 Wrocław, Poland}

\author{Sujan Maity}
\affiliation{School of Physical Sciences, Indian Association for the Cultivation of Science, 2A \& B
Raja S. C. Mullick Road, Jadavpur, Kolkata - 700032, India
}

\author{Somsubhra Ghosh}
\affiliation{School of Physical Sciences, Indian Association for the Cultivation of Science, 2A \& B
Raja S. C. Mullick Road, Jadavpur, Kolkata - 700032, India
}

\author{Bikash Das}
\affiliation{School of Physical Sciences, Indian Association for the Cultivation of Science, 2A \& B
Raja S. C. Mullick Road, Jadavpur, Kolkata - 700032, India
}

\author{Koushik Dey}
\affiliation{School of Physical Sciences, Indian Association for the Cultivation of Science, 2A \& B
Raja S. C. Mullick Road, Jadavpur, Kolkata - 700032, India
}

\author{Satyabrata Bera}
\affiliation{School of Physical Sciences, Indian Association for the Cultivation of Science, 2A \& B
Raja S. C. Mullick Road, Jadavpur, Kolkata - 700032, India
}


\author{Suresh Bhardwaj}
\affiliation{UGC-DAE Consortium for Scientific Research, University Campus, Khandwa Road, Indore-452001,India}

\author{Shibabrata Nandi}
\affiliation{Forschungszentrum J\"{u}lich GmbH, J\"{u}lich Centre for Neutron Science (JCNS-2) and Peter Gr\"{u}nberg Institut (PGI-4), JARA-FIT, 52425 J\"{u}lich, Germany}
\affiliation{RWTH Aachen, Lehrstuhl f\"{u}r Experimentalphysik IVc, J\"{u}lich-Aachen Research Alliance (JARA-FIT), 52074 Aachen, Germany}

\author{Subhadeep Datta*}
\affiliation{School of Physical Sciences, Indian Association for the Cultivation of Science, 2A \& B
Raja S. C. Mullick Road, Jadavpur, Kolkata - 700032, India
}

\begin{abstract}

We report anisotropic magnetodielectric (MD) coupling in layered van der Waals (vdW) antiferromagnetic (AFM) FePS$_3$ (N\'eel temperature $T_{\mathrm{N}}$ $\sim$ 120K) with perpendicular anisotropy. Above $T_N$, while dielectric response function along \textit{c}-axis shows frequency dependent relaxations, in-plane data is frequency independent and reveals a deviation from phonon-anharmonicity in the ordered state, thereby implying a connection to spin-phonon coupling known to be indicative of onset of magnetic ordering. At low temperature (below 40 K), atypical anomaly in the dielectric constant is corroborated with temperature dependent DC and AC susceptibility. The magnetodielectric response across this anomaly differs significantly for both, in-plane and out-of-plane cases. We have explained this in terms of preferential orientation of magnetic AFM-z alignment, implied by the in-plane structural anisotropy as confirmed by \textit{ab-initio} calculations. Controlling relative strength of magnetodielectric coupling with magnetic anisotropy opens up a strategy for tracking subtle modifications of structure, such as in-plane anisotropy, with potential application to spintronic technologies.

\end{abstract}


\maketitle

\section{Introduction}

Multifunctional devices based on spin-charge coupling involve low-frequency shift of dielectric constant with magnetic ordering \cite{kcfrcf}. Additionally, presence of the magnetic anisotropy (MA) may drive the exotic spin textures and, in turn, lead to electric field control of the magnetic ground state \cite{lawes,spinflop}. Two-dimentional (2D) vdW magnetic materials is of particular interest due to the presence of MA originating from the interaction between the magnetic moments and the crystal field. Also, these materials indicate high degree of stability in the long-range spin order and may be described using suitable spin-Hamiltonian of Heisenberg-, XY- or Ising-type. Moreover, recent reports suggest the effective interactions between magnetization and electric polarization in 2D magnets \cite{rucl3,jiang,mnps}. Exploring new routes to MD coupling, such as complex spin structures, magnetostructural, and magnetoelastic effects, has become important from fundamental point of view as well as device applications \cite{hill,catalan,dd,scott,triangularlatt,kimura,thinfilm1,thinfilm2}. 

Other than the charge/spin transport measurements, the coexistence of electric and magnetic orders in a few-layer AFM can be detected from phonon anomalies \textit{via} $\mu$-Raman spectroscopy or optical second harmonic generation. However, direct probing of dielectric constant with varying temperature, frequency and magnetic field  parameters in transition-metal (M) trichalcogenides (MPX$_3$, X = S, Se), in their bulk forms, is still largely missing from the literature. The vdW gaps in relatively air-stable MPX$_3$ ($\sim$ 2 - 3 \AA) host  interstitial sites that have shown to facilitate intercalation of guest ions  \cite{intercalation1,intercalation2} and can thus provide hopping sites in the \enquote{out-of-plane} direction. This is absent in the \enquote{in-plane} direction since it is constituted of strong covalent bonds. In the conventional \enquote{parallel-plate-capacitor} measurement scheme, anisotropic lattice and spin texture in these magnetic insulators result in contrasting dielectric properties with different charge carrier transport mechanism in the \enquote{in-plane} (\textbf{E} $\parallel$ \textit{c-axis}) and \enquote{out-of-plane} (\textbf{E} $\perp$ \textit{c-axis}) directions.

Here, we present a comprehensive low temperature dielectric spectroscopy of a layered antiferromagnet FePS$_3$ with $T_{\mathrm{N}}$ $\thicksim$ 120 K. The dielectric function measured along the \textit{c-axis} is frequency independent in the AFM phase, but shows the onset of dielectric relaxations above $T_{\mathrm{N}}$. On the other hand, the \enquote{in-plane} function, remains frequency independent throughout and shows deviation from the usual anharmonic behaviour at $T_{\mathrm{N}}$ which can be correlated to the spin-phonon coupling from our previous study \cite{datta}. The \enquote{out-of-plane} relaxations have been corroborated with temperature dependent dc conductivity and analysed in terms of the small polaron hopping model. A distinct anomaly is observed in the dielectric constant around 50 K and is also reflected in AC magnetic susceptibility. These have been explained in terms of preferential orientation of the AFM-z phase alignment within the plane, enabled by the in-plane structural anisotropy, facilitated by distortion of lattice parameters at low temperatures and is supported by theoretical considerations. A contrasting phenomenon is observed in the magneto-dielectric response across this anomaly between the \enquote{out-of-plane} and \enquote{in-plane} directions with spin-phonon correlation assisted magneto-dielectric coupling showing up for the \enquote{in-plane} case.



\section{Sample preparation and measurement}
\label{subsec:exp}

Single crystals of FePS$_3$ were grown by the chemical vapor transport method, characterized and studied \textit{via} X-ray diffraction, energy dispersive X-Ray analysis, DC and AC susceptibility. Low temperature dielectric spectroscopy with varying frequency and magnetic field were performed following the \enquote{parallel-plate} geometry for the in-plane and out-of-plane measurement on exfoliated bulk material (\fig{schem}(a)). For computational studies, the static dielectric properties were calculated by means of density functional perturbation theory implemented in the VASP software. Details of the crystal growth, measurement schemes and the computational studies are given in the supplementary information \cite{suppli}.

\section{Results and discussion}
\label{sec:results_discussions}


\begin{center}
\bfseries{A. Dielectric Spectroscopy}\\
\end{center}

\begin{center}
\bfseries{(i) Region around the N\'eel temperature:}\\
\end{center}

The out-of-plane (\textbf{E}$\|c$) dielectric constant ($\varepsilon'$) of FePS$_3$ as a function of temperature for various frequencies is shown in \fig{attn}(a). In the AFM phase, below $T_{\mathrm{N}}$, $\varepsilon'$ is almost frequency and temperature independent, representing the static part of dielectric constant due to the electronic and ionic contributions \cite{kao}. 

As the temperatures is increased, in the paramagnetic (PM) phase, a rapid increase in $\varepsilon'$ is observed with the onset of frequency dependent dielectric relaxations. This also manifests as peaks in the loss factor $\tan \delta$ (not shown here) which shows wide shifts towards higher temperature with increasing frequency indicating thermally activated relaxation mechanism \cite{chanda}. Two different types of relaxations in the PM state for the given temperature window can be identified, as marked by A and B in \fig{attn}(a). In the frequency range being probed, the relaxations can either arise from Debye/Debye-like relaxation or from the charge accumulation near boundaries, otherwise called Maxwell-Wagner (MW) relaxations \cite{dd,kao}. The slope calculated from the $\log(\varepsilon''$) vs $\log(f)$ plot is found to be (-1) in the B region  (see SI \cite{suppli}) suggesting the presence of the MW relaxation \cite{dd}. For a vdW material like FePS$_3$, the constituent layers in the bulk  along the out-of-plane direction is separated by vdW gaps, which may lead to interfacial charge accumulation between layers. 

MW relaxation model, however, fails to fit the data in region A [see \fig{attn}(a) and SI \cite{suppli}]. This interim temperature regime (region A) was fitted with Debye-like model (see SI \cite{suppli}) with a characteristic relaxation time and can be attributed to response of the polar microregions in field \textbf{E}. Accordingly, combined MW and Debye-like model explains the data over the entire temperature regime in A and B. For Debye-like relaxation, the relaxation time ($\tau_0$) and activation energy (\~E) determined from the Arrhenius relation (see SI \cite{suppli}) were found to be 1.5 $\times$ 10$^{-7}$ s and 219 meV, respectively. The large relaxation time indicates a hopping type conduction of quasi particle like small polarons (SP) through interstitial sites in vdW gaps \cite{kao,mott}. Considering the nearest-neighbour (NN) SP hopping, the temperature dependent DC resistivity ($\rho_{dc}$/T versus 1000/T plot in \fig{attn}(b)) measured in the top-bottom configuration, can be described by \cite{chen,mott}:

\begin{equation}
\rho = CT\exp{\left(\frac{E_A}{k_BT}\right)}
\end{equation}

where E$_A$ is the activation energy and k$_B$ is Boltzmann’s constant and C is the prefactor. The activation energy calculated from the fit is 170 meV which corroborates with that calculated from Arrhenius relation. The NN-small polaron model (Eq. 1) fits well with the data for temperatures above 180 K but shows deviation below 180 K. Alternative mechanisms like Mott's variable range hopping (VRH) or Shklovskii-Efros variable range hopping (ES-VRH) fail to fit the data in the temperature regime below 180 K [see inset of \fig{attn}(b)] \cite{giri,rgo}. Considering the limiting case approximation, where SPs can penetrate to neighboring sites by phonon-induced tunneling effect, the hopping-type transport becomes dominant for $T>0.5\hbar\omega/k_B$ \cite{holstein,kao}, where $\omega$ is the optical mode angular frequency. Below $T\approx$ 180 K lies the non-Arrhenius regime, dominated by tunnelling transport of polarons, which puts a figure on vibrational spin-phonon coupled Raman-active bulk mode ($\omega$) at 250 $cm^{-1}$, reported in our previous study \cite{datta}.






The \enquote{in-plane} dielectric constant shows no frequency dependent dielectric relaxations for the entire temperature and frequency range (see \fig{attn}(c)), which asserts the effect of vdW gaps in the \enquote{out-of-plane} direction. One may note that the samples used in this study are pristine bulk flakes and \enquote{well-stamped} \textit{via} micro-manipulation technique.  
The temperature-dependent low-frequency dielectric permittivity ($\varepsilon_{0}$(T)) of an insulator without any structural, ferroelectric, or magnetic phase transitions is characterized by Einstein-type function as \cite{kcfrcf,seehra}:

\begin{equation}
\varepsilon_0(T) = \varepsilon_0(0) + \frac{A}{\exp{\frac{\hbar\omega^*}{k_BT}}-1}
\end{equation}

where $\varepsilon_{0}(0)$ and A are constants and $\omega^*$ is the frequency of the effective infrared (IR) active optical phonon with a dominant dielectric strength at zero temperature.  A frequency value of $\approx$ 431 cm$^{-1}$ has been predicted as a strong IR active mode in bulk FePS$_3$ in an earlier report by Joy \textit{et al.}  \cite{infrared} and is thus chosen as $\omega^*$ which fits well with the experimental data for $f$ = 120 Hz (inset \fig{attn}(c)). The \enquote{in-plane} dielectric data deviates from the anharmonic fit around $T_{\mathrm{N}}$, $\approx$ 120 K, similar to Raman spectroscopic studies \cite{datta}, indicating the influence of spin-phonon coupling in FePS$_3$. 
The temperature variation of normalized in-plane AC resistance ($R$/R$_{max}$) show three different regions marked as A, B \& C (see inset \fig{attn}(d)). At lower temperatures, below 120 K (region A), temperature independent resistance for all frequencies can be observed. At higher temperatures, a sudden drop in resistance has been recorded for all the measured frequencies (region B). However, the drop in resistance have been found to start from higher temperatures for higher frequencies. Inset (ii) of \fig{attn}(d) shows ln R \textit{vs.} 1000/T plot from 250 K to 300 K (region C) which agrees well with the Arrhenius law  $R$ $\sim$ exp(E$_a$/2k$_B$T) with the activation energy (E$_a$) as 81.6 meV \cite{Sun}. However, below 250 K, a pronounced upturn in the resistivity data is clearly seen from where the thermally activated Arrhenius model fails to explain the temperature variation of $R$. This upturn behavior can be explained by spin-charge scattering using the relation, $\rho$ = $A$ + $B$ ln(T$_{SF}$/$T$), where A \& B are constants and T$_{SF}$ is the temperature below which spin fluctuation starts \cite{Dgong}. Here, we incorporate the concept of spin fluctuations and spin charge scattering to explain the resistivity upturn as there are evidences of spin dynamics and magnon polarons in this compound \cite{xzhang, mainak}. Interestingly, in this temperature range, frequency dependent dielectric relaxation is also prominent (Fig. 2(a)). The T$_{SF}$ obtained from the fitting (inset (iii) in \fig{attn}(d)) is 213 K, associated with the onset of spin fluctuation. Below this temperature range, $R$/R$_{max}$ are found to increase slowly which indicates the suppression of spin-charge scattering. Also, several magnetic and dielectric anomalies can be observed to present which do not influence much on temperature variation of resistivity \cite{simonson}.

\begin{center}
\bfseries{(ii) Region around 50 K:}\\
\end{center}

\begin{figure}[h!]
\centerline{\includegraphics[scale=0.5, clip]{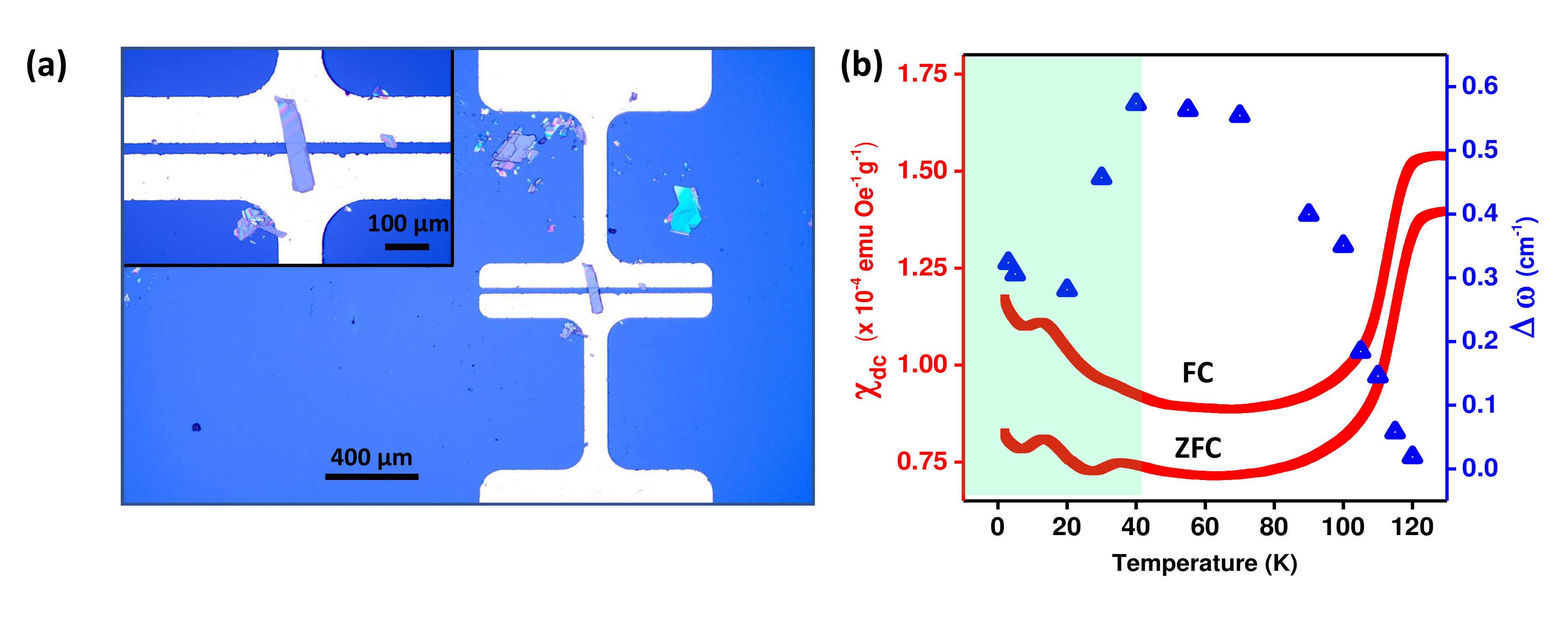}}
\caption{Iron phosphorus trisulfide (FePS$_3$): (a) Optical image showing two-probes with a FePS$_3$ flake stamped on it. Inset shows zoomed-in image of the flake. (b) Low-temperature portion (at and below $T_{\mathrm{N}}$) of dc susceptibility taken at a field of 500 Oe plotted against temperature on the left axis. Right axis shows the deviation from anharmonicity ($\Delta\omega$) for the  Raman peak at 285 cm$^{-1}$ plotted as a function of temperature. The former shows a distinct upturn below 40 K much below N\'eel temperature $\thicksim$ 120 K at which antiferromagnetic ground state is established. The latter shows a decrease in $\Delta\omega$ around 40 K. 
\label{schem}}
\end{figure}

\begin{figure}[ht]
\centerline{\includegraphics[scale=0.5, clip]{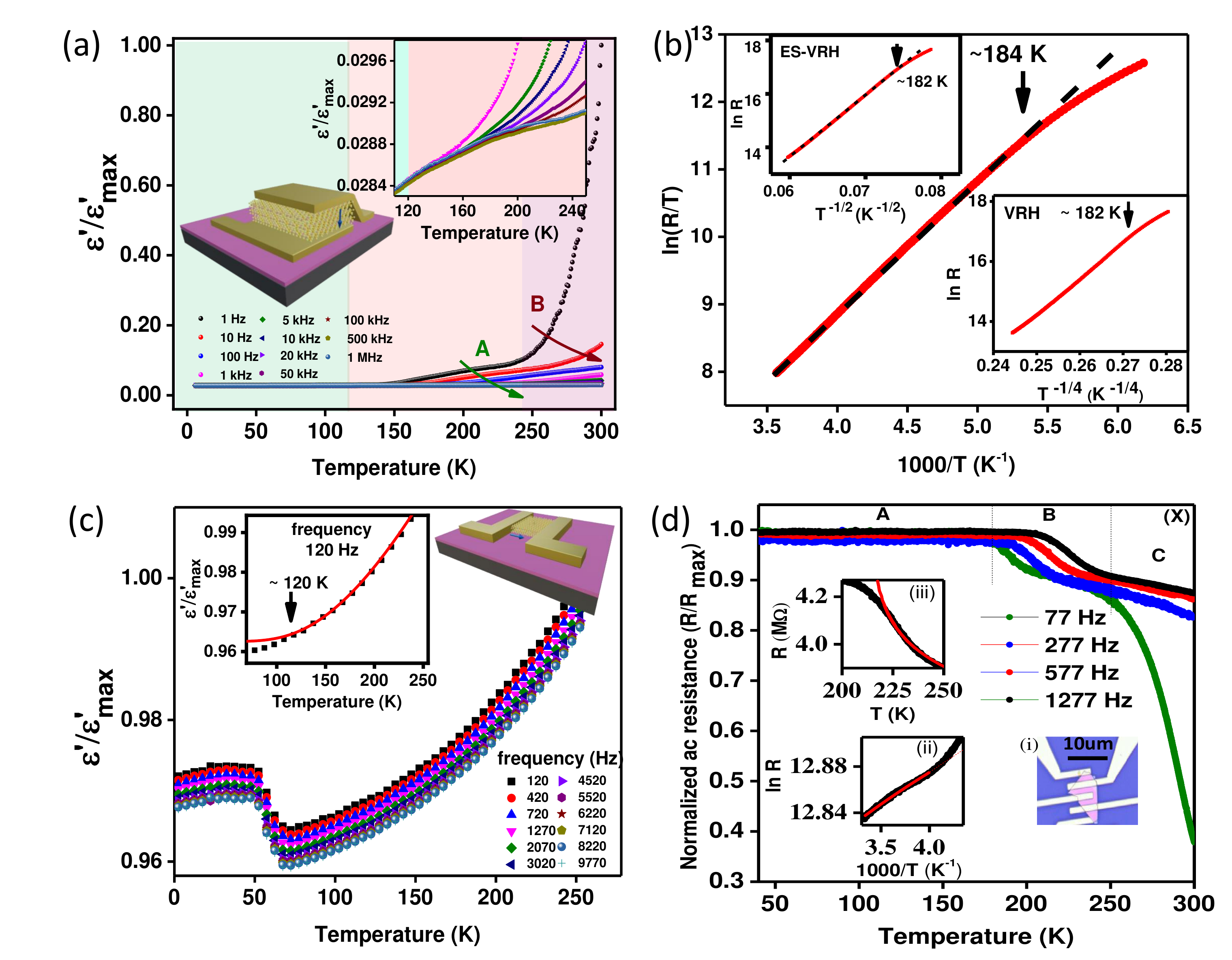}}
\caption{Dielectric spectroscopy: (a) Temperature dependent (3 - 300 K) measurement at different frequencies for out-of-plane geometry.  Two different relaxation regions are marked as A and B. Inset shows the dispersion for higher frequencies in the intermediate temperature regime. (b) Arrhenius plot for out-of-plane resistance showing deviation at $\thicksim$ 184 K. Similar deviation is seen in ES-VRH and VRH shown in top and bottom inset. (c) Temperature dependent (3 - 250 K) measurement at different frequencies for in-plane geometry.  Inset shows normalised dielectric permittivity with Einstein fit showing deviation around 120 K. (d) Temperature dependent normalized in-plane AC resistance of FePS$_3$ bulk flake (see inset (i)) with fitting of resistivity in two different temperature ranges (250 K - 300 K in the inset (ii) and 220 K - 250 K and the inset (iii)). 
\label{attn}}
\end{figure}

\begin{figure}[h!]
\centerline{\includegraphics[scale=0.6, clip]{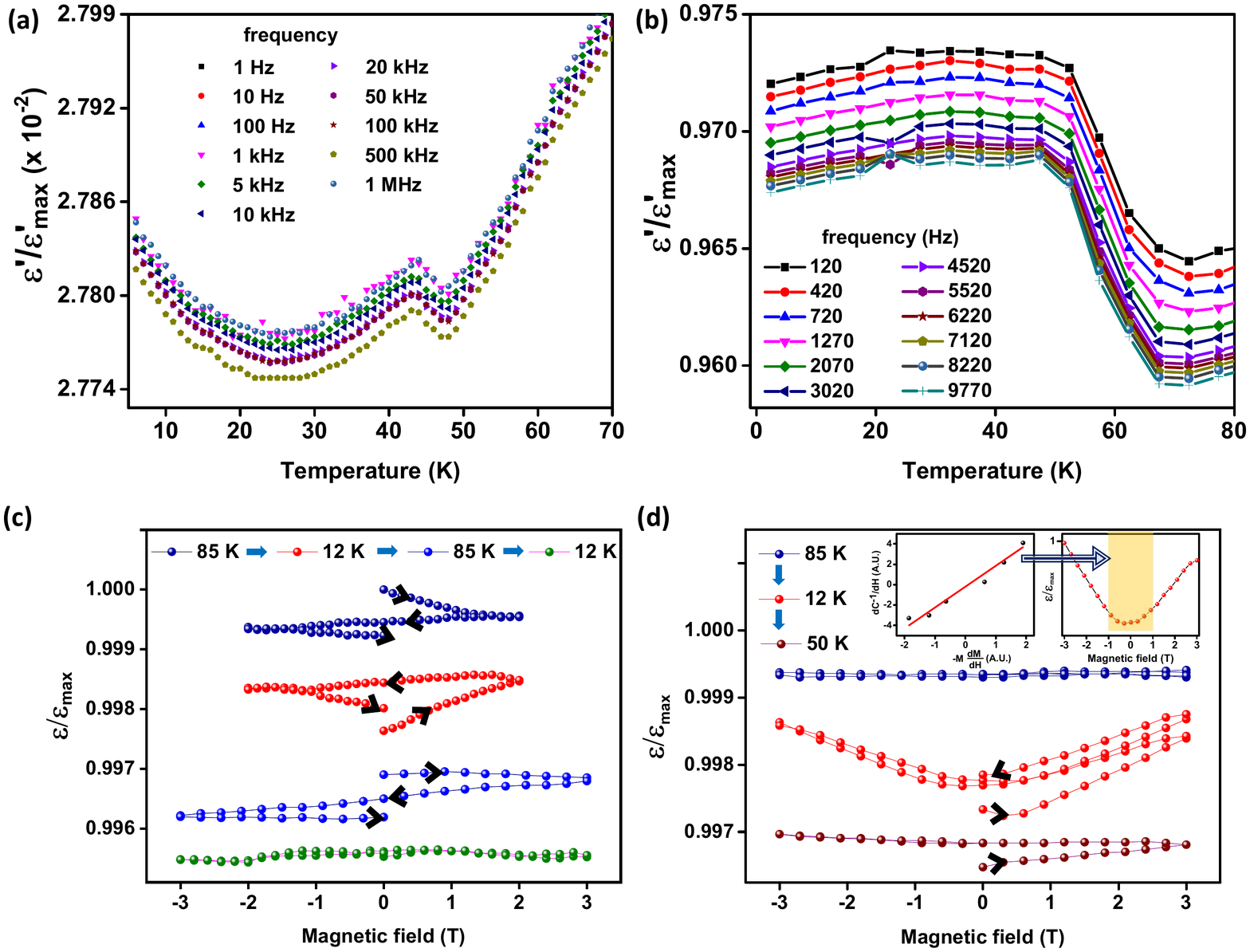}}
\caption{(a) Magnetodielectric coupling: (a) \& (b) Dielectric spectra in the low temperature region, (a) Out-of-plane measurements for different frequencies showing a kink around 50 K followed by a change in the nature of the curve below 40 K. (b) In-plane measurement showing a prominent jump at temperature around 50 K. (c) \& (d) Dielectric response as a function of magnetic field sweep, below and above the dielectric anomaly $\thicksim$ 50 K. (c) Measurements with out-of-plane geometry were taken at selected temperatures in the following order: 85 K, 12 K, 85 K, 12 K. The nature of response on virgin sample shows distinct difference between 85 K and 12 K, as shown by the navy and red plots. Even though the 85 K data reproduces, the response at 12 K is lost (see Section A(iii)). (d) In-plane measurement has been taken consecutively at 85 K (navy), 12 K (red), 50 K (brown) which show no locking effect and a prominent magnetodielectric coupling is observed at 12 K. 
\label{lowtn}}
\end{figure}

A close inspection of the low temperature dielectric data reveals a frequency independent anomaly in $\varepsilon'$ around $\thicksim$ 50 K reflected as a sudden jump in both, \enquote{out-of-plane} (see \fig{lowtn}(a)) and \enquote{in-plane} (see \fig{lowtn}(b)) geometries. It is noteworthy that the characteristic Raman modes unveils an unusual downturn at $\thicksim$ 40 K in the deviation from phonon anharmonicity ($\Delta\omega$), where $\Delta\omega$ a signature of the strength spin-phonon coupling arising at $T_{\mathrm{N}}$ [see \fig{schem}(b)] \cite{datta}. Moreover, the magnetization data reflects similar anomaly where $\chi_{dc}$ shows an upturn from the AFM ground state below T $<$ 40 K [\fig{schem}(b)]. Anomaly in $\varepsilon'$ is usually correlated to magnetic phase transition \cite{kimura,lawes,park} or ferroelectric ordering \cite{schrettle1,schrettle2,shi}. Note that magnetic field induced quantum fluctuation in AFM at low temperature can also trigger such anomaly but does not match well with the scale ($\Delta \varepsilon$) (discussed in SI \cite{suppli}). However, a displacive-type ferroelectric transition, especially in the out-of-plane case, might be a possibility \cite{reviewferroelec}.    





\begin{center}
\bfseries{(iii) Magnetodielectric response:}\\
\end{center}


We demonstrate anisotropic magnetodielectric response in FePS$_3$ in magnetic field space (\textbf{H}) applied parallel to the c-axis in FePS$_3$ for both, \enquote{in-plane} and \enquote{out-of-plane} configurations. For the \enquote{out-of-plane} case, the manner in which applied magnetic fields change the dielectric response differs significantly above and below the dielectric anomaly seen around 50 K. \fig{lowtn}(c) shows the change in dielectric permittivity when magnetic field is gradually sweeped between 0 T and $\pm$ 2 T. The frequency is set to 100 kHz such that space–charge artifacts, contributing to magnetodielectricity, can be avoided. The first measurement taken at 85 K (navy) shows continuous decrease in permittivity with increasing and subsequent decrease in magnetic field. There is a change of $\thicksim$ -0.08\% between initial and final value after one complete cycle. Next, the temperature is lowered to 12 K and another cycle is taken (red). There is a marked change in the nature of dielectric response wherein the permittivity initially increases rapidly when field changes from 0 T to 2 T but thereafter decreases from 2 T to -2 T and continues to decrease from -2 T to 0 T. The cycle is hysteric and the maximum change in capacitance is $\thicksim$ +0.08\%. Next, the temperature is increased back to 85 K where the initial nature of the curve is reproduced even with increase in magnetic field to $\pm$ 3 T. However, when the cycle is subsequently repeated at 12 K, the change in permittivity is now negligible ($\thicksim$ 0.009\%). The curve loops onto itself with increasing and decreasing field and the initial behaviour, seen in the virgin sample, is now lost. Even though the exact phenomenon behind such distinctive difference in the magnetodielectric responses at 85 K and 12 K require further studies, the observation demonstrate that the magnetic field induces irreversible change in dielectric permittivity at low temperature. At lower temperature ($\thicksim$ below 50 K), the temperature induced structural frustration causes the micro-polar regions to align differently than that at 85 K, such that the application of magnetic field causes locking of the moments which do not return to original state even under a demagnetising field.

For the \enquote{in-plane} case (\fig{lowtn}(d)) taken at 100 kHz, however, the magnetodielectric response is significantly different than \enquote{out-of-plane} case. There is no permanent locking effect and the measurements taken consecutively at 85 K (navy), 12 K (red), 50 K (brown) and 12 K (not shown) show a distinct magnetodielectric coupling which is most prominent at 12 K ($\thicksim$ +0.14\%) and decreases with temperature becoming negligible at 85 K. This can possibly be attributed to the spin-phonon coupling \cite{kcfrcf} observed in our previous report \cite{datta}. 

From Landau free energy considerations, the variation of inverse dielectric susceptibility function (which scales with inverse capacitance) can be expressed as \cite{jfscott}:

\begin{equation}
\frac{d\chi^{-1}}{dH} = \sum_{i,j,k=0}^{\infty}  D(i,j,k) i (i-1) P^{i-2} j \frac{dM}{dH} M^{j-1} \epsilon^k
\end{equation}

where $\chi$ is the dielectric susceptibility, D(i,j,k) is a constant, P is the electric polarization, M is the magnetization, H magnetic field and the $\epsilon$ is the strain.

Careful examination of the derivative of the inverse dielectric susceptibility with magnetic field, can give information about the coupling terms in the Landau free energy expansion.
The P$^2$ M$^2$ coupling, which is always allowed by symmetry, if present in the material, then $\frac{d(1/C)}{dH}$ should be proportional to $M(\frac{dM}{dH})$ \cite{jfscott}. First inset of \fig{lowtn}(d) shows plot of $\frac{d(1/C)}{dH}$ vs $-M(\frac{dM}{dH})$ at 12 K which gives a straight line between $\pm$ 1 T (second inset \fig{lowtn}(d)) but deviates thereafter.




With lowering of temperature, the distortion in lattice parameters (with the length of the a- and b-axis decreasing and increasing, respectively \cite{murayama,jernberg}), coupled with anisotropy, results in complex interaction within the domains leading to frustration in the system below $\thicksim$ 50 K and subsequent freezing. This explains the large temperature shift in $\chi'$ peaks and the anomalous jump in the dielectric spectra. This might lead to domain wall motion or related dynamics at low temperatures which would be governed by the anisotropy constants \cite{nauman}. Anomalous nature of $\chi'$, showing two sets of frequency dependent peaks, may point towards more than one domain wall related phenomenon. Such temperature-induced domain wall movement has also been observed in other Ising-systems like CoNb$_2$O$_6$ \cite{domainising}. Note that there has been theoretical predictions on the magnetic field and electrical current controlled domain wall dynamics in 2D vdW magnets like CrI$_3$, CrBr$_3$ and MnPS$_3$ \cite{wahab,mnps3domain}. 

\section{Theoretical results}
In order to understand the magnetodielectric measurements, we carried out the \textit{ab initio} calculations of bulk FePS$_3$ system. The magnetic ions (Fe) are arranged within the honeycomb lattice and exhibit antiferromagnetic zig-zag (AFM-z) ordering. A previous temperature-dependent X-ray diffraction (XRD) study reported that the in-plane lattice constant ratio deviating from the hexagonal symmetry \cite{murayama}. The latest XRD measurements demonstrated the nonequivalent Fe-S bond lengths within the FeS$_6$ octahedron, pointing to the existence of crystallographic in-plane anisotropy \cite{Ellenore}. This might be a consequence of the symmetry breaking of the honeycomb structure with a further adjustment of the nearest neighbor distance between the Fe atoms \cite{amirabbasi}, implying a preferred direction of the AFM-z phase within the monolayer plane \cite{Ellenore}. 

\begin{figure}[h!]
\includegraphics[scale=0.5, clip]{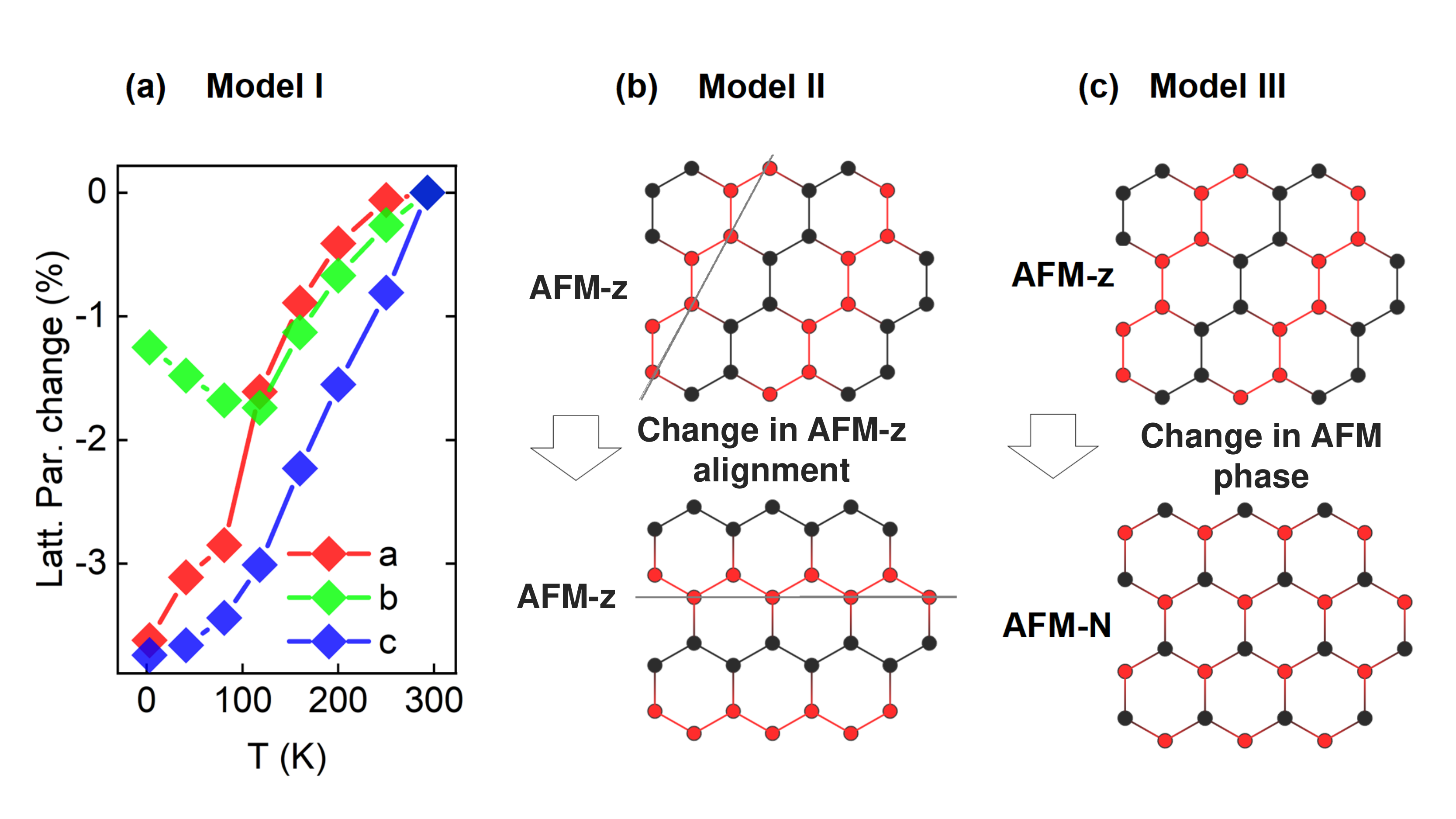}
\caption{The impact of three different factors on the dielectric properties are examined. In Model I various lattice parameters are adopted from the XRD measurements [Ref. \cite{murayama}]. In Model II and Model III, the change of the AFM-z orientation and two lowest magnetic phases are assumed, respectively.
\label{AFMmodels}}
\end{figure}

To elucidate the origin of the prominent jump around 50 K for the in-plane geometry (see \fig{lowtn}(b)), we examine three plausible factors that might affect the dielectric properties of the bulk materials as presented in \fig{AFMmodels}. Namely, we examine the change of the lattice parameters in respect to elevating temperatures (\textbf{model I}), the change of the zigzag orientation within the monolayer frame (\textbf{model II}), and the impact of the magnetic phase (AFM-z, AFM-N\'{e}el) (\textbf{model III}). For all of these approaches, we examine the in-plane ($\varepsilon_{\|}$) and out-plane ($\varepsilon_{\bot}$) contributions of dielectric constant $\varepsilon_0$. Note, that the $\varepsilon_0$ represents a macroscopic static response containing both the ionic ($\varepsilon_{ion}$) and the electronic response ($\varepsilon_{\infty}$) \cite{suppli}. The stronger polarization is expected for the covalent bonds (in-plane ones) and weaker for the vdW type bonding. Since the in-plane contributions are around 5 times larger than out-of plane ones (see Fig. S6 and Table 1 in SI), and the prominent jump is observed for the in-plane geometry, we only discuss the in-plane dielectric contributions below (for the details of out-of plane contributions, see SI \cite{suppli}). To compare the theoretical results with experimental values we define the relative dielectric constants as $\delta\varepsilon_{\|}$ =($\varepsilon_{\|} - \varepsilon_{ref}$)/$\varepsilon_{ref}$, where $\varepsilon_{\|}$ and  $\varepsilon_{ref}$, are particular and reference values of dielectric constant, respectively. The reference value is taken as a minimal value within the  range under consideration. Now, we briefly explain each of the models.

As reported previously by XRD studies, the a/b lattice ratio exhibits strong temperature dependence \cite{murayama}. In \textbf{model I}, we assumed the lattice parameter changes reported by Murayama et al. \cite{murayama}.  Note, that the temperature was not included explicitly in our calculations, and reflects only the lattice parameters measurement’s taken from 4 K up 300 K \cite{murayama}. Since the in-plane structural anisotropy was recently reported \cite{Ellenore}, the AFM-z phase exhibits preferred alignment within the layer, and its change might impact the dielectric properties. Thus, in \textbf{model II} we employ the change of the orientation of AFM-z phase within the monolayer frame (see Fig. S7). In \textbf{model III}, we consider two lowest magnetic phases: AFM-z and AFM-N \cite{magda1}, assuming the lattice parameters extracted from experimental measurements around the kink ($\sim$ 50 K). The results of all three models are collected in Table 1. 

In \textbf{model I} the changes of the in-plane dielectric contributions are small upon the changes of the lattice parameters. Albeit, there is a visible kink in ionic contribution at 80 K for U=5.3 eV (see  Fig. S7 (c), (d)), however  it is not shown for other Hubbard U parameter (see S7 (a), (c) for U=2.6 eV). In \textbf{model II} the change in the alignment of AFM-z order implies a larger increase of the ionic relative dielectric contribution $\delta\varepsilon_{ion}$ (0.6 \%) than compared to \textbf{model I} (0.3\%-0.5\%). The strongest changes of in-plane dielectric properties (around 8\%, see Table) are exposed by the change of the magnetic phases, AFM-z and AFM-N ones. In particular, the largest values of dielectric in-plane constants are obtained for AFM-N phase. In addition, our results reveal that the magnetic ground state (AFM-z) is robust against the employed range of lattice parameters (Fig. S7(b)), in line with recent theoretical reports for the other MPX$_3$ antiferromagnetic structures \cite{magda1}.  Although, the relative change of the magnetic phase is plausible to be observed in higher temperature ($>$ 50K), as indicated by our DFT+U results (see explanation in SI), no significant kinks, jumps, changes of the in-plane dielectric properties are visible for the Ne\'el temperature at 120 K. In addition, the relative changes of the in-plane contributions around (3-5\%), are rather large in comparison to the experimentally observed ones (0.8\%-1\%). Hence, the \textbf{model III} can be excluded as being origin of the jump around 50 K. On the other hand, the structural in-plane anisotropy reported recently \cite{Ellenore}, and the changes in lattice parameters \cite{murayama} imposing a preferred orientation of the magnetic alignment. The energy difference of 5.6 meV per magnetic ion reported in [\cite{Ellenore}, see SI therein], indicate that the thermal energy could rotate the AFM-z alignment at the temperature of around 65 K.  The relative change in magnetic alignment impose the change of the in-plane dielectric contributions equal to $\delta\varepsilon_0=0.2\%$ (model II), which is in the same order as observed experimentally ($\delta\varepsilon_0=0.8\%$). Note that, the theoretical value obtained within \textbf{model II} could be further enhanced by including the relative changes of the lattice parameters, as indicated by the \textbf{model I}. Thus, the prominent jump visible around 50 K for the in-plane measurements of the dielectric permittivity might be attributed to the change of the orientation of the AFM-z phase alignment within the plane, enabled by the in-plane structural anisotropy.

\begin{table}[ht]
\caption{In-plane contribution of the relative dielectric constants defined as $\delta\varepsilon_{\|} =(\varepsilon_{\|} - \varepsilon_{ref})/\varepsilon_{ref}$. In particular, for Model I, a  $\varepsilon_{ref}$ is taken as a minimal value from the range 3 K-81 K. Regarding, the total contribution of  $\delta\varepsilon_0$, each value of $\varepsilon_{\|}$, $\varepsilon_{\|}^{ref}$  is the sum of the ionic and electronic contributions, and thus, $\delta\varepsilon_0$ is not a sum of $\delta\varepsilon_{\infty}$ + $\delta\varepsilon_{ion}$. In the last row, the energy difference $\vartriangle$E and its corresponding thermal energy is presented.  In the case of Model I the $\vartriangle$E is evaluated for the magnetic ground state (AFM-z), in model II the 5.6 meV is taken from the Ref. \cite{Ellenore} (see SI therein), and in model III the $\vartriangle$E is between two magnetic phases AFM-z and AFM-N (see Fig. S7(b))}

\label{tab:dielectric}
\centering
\resizebox{\textwidth}{!}{\begin{tabular}{|c|c|c|c|}
\hline
\textit{In-plane dielectric} & \textbf{Model I}: & \textbf{Model II}: & \textbf{Model III}: 
\tabularnewline
\textit{contribution $\delta\varepsilon_{\parallel}$}[\%] & (change of the & (change of the AFM-z & (change of the 
\tabularnewline
& lattice parameters) & alignment within  layer) & magnetic phase) 
\tabularnewline
 \hline
Ionic $\delta\varepsilon_{ion}$ & 0.3\% (U=5.3 eV)  & 0.6\% (U=5.3 eV) & 7.9\% (U=5.3 eV)   
\tabularnewline
 & 0.5\% (U=2.6 eV) & {} & 7.6\% (U=2.6 eV) 
  \tabularnewline
\hline
 Electronic $\delta\varepsilon_{\infty}$ & 0.12\% (U = 5.3 eV)  & 0.01\% & 7.9\% (U=5.3 eV)   
\tabularnewline
 & 0.14\% (U=2.6 eV) & {} & 0.9\% (U=2.6 eV) 
 \tabularnewline
\hline  
  Total $\delta\varepsilon_0$ & 0.2\% (U=5.3 eV)  & 0.2\% & 5.2\% (U=5.3 eV)   
\tabularnewline
 & 0.1\% (U=2.6 eV) & {} & 3.3\% (U=2.6 eV) \\
\hline
 
\multicolumn{4}{|c|}{ Experimental:{   } $\sim$ 0.8\% (see Fig. 3(b), obtained for range (3 K - 70 K))}\\
\hline
 
$\vartriangle$E[meV per magnetic atom] & 1.6 meV & 5.6 meV [\cite{Ellenore}] & 11.5 meV for U=5.3 eV,  
\tabularnewline
(thermal energy) & (19 K) & (65 K) & (133 K)
\tabularnewline
& & & 3.9 meV  for U=2.6 eV, 
\tabularnewline
& & & (45 K)\\
\hline 
\end{tabular}}
\end{table}

\section{Conclusions}
\label{sec:conclusion}

To summarize, we examined the magnetodielectric properties of FePS$_3$ which shows anisotropic behaviour in the \enquote{in-plane} and \enquote{out-of-plane} direction which can be attributed to contrasting nature of bonding and spin texture in these two geometries. A prominent anomaly in the AFM phase ($\thicksim$ 50 K) is observed in the dielectric spectra, supported by AC susceptibility measurements, has been explained in terms of complex interaction in the domains, which might, in turn, lead to domain wall movements. Computationally, three plausible models have been examined. Structural in-plane anisotropy along with the non-equivalent changes in lattice parameters imposes a preferred orientation of the magnetic alignment leading to a kink in dielectric constant at low temperatures. Tailoring the structural anisotropy in 2D magnets by tuning magnetodielectric coupling may be promising for future spin-logic device applications. 



\begin{acknowledgments}
We are grateful to CSS facility at IACS and Prof. S. Giri for the support in the dielectric measurements. The authors acknowledge fruitful discussion with Prof. K. Sengupta, Dr. M. Mondal, Dr. K. D. M. Rao, Dr. M. Palit, Mr Soumik Das, and Ms. S. Baidya . AG would like to thank Dr. Anupam Banerjee and Shameek Mukherjee. M.B. acknowledges support by the University of Warsaw within
the project “Excellence Initiative-Research University” programme. Access to computing facilities of PL-Grid Polish Infrastructure for Supporting Computational Science in the European Research Space and of the Interdisciplinary Center of Modeling (ICM), University of Warsaw are gratefully acknowledged. SM is grateful to DST-INSPIRE for his fellowship. SG acknowledges CSIR for the fellowship (File No. 09/080(1133)/2019-EMR-I). SD would like to acknowledge DST-SERB grant No. CRG/2021/004334 and e-beam lithography facility of TRC at IACS. The authors are also thankful to the facilities at UGC-DAE-CSR-Indore.
\end{acknowledgments}


\begin{thebibliography}{99}

\bibitem{kcfrcf}
R. Dubrovin, N. Siverin, P. Syrnikov, N. Novikova, K. Boldyrev, and R. Pisarev,  Lattice dynamics and microscopic mechanisms of the spontaneous magnetodielectric effect in the antiferromagnetic fluoroperovskites KCoF$_3$ and RbCoF$_3$. {\em Phys. Rev. B}. \textbf{100}, 024429 (2019)

\bibitem{lawes}
G. Lawes, A. Ramirez, C. Varma, and M. Subramanian,  Magnetodielectric effects from spin fluctuations in isostructural ferromagnetic and antiferromagnetic systems. {\em Phys. Rev. Lett}. \textbf{91}, 257208 (2003)

\bibitem{spinflop}
T. Kolodiazhnyi, H. Sakurai, and N. Vittayakorn,  Spin-flop driven magneto-dielectric effect in Co$_4$Nb$_2$O$_9$. {\em Applied Physics Letters}. \textbf{99}, 132906 (2011)

\bibitem{rucl3}
T. Aoyama, Y. Hasegawa, S. Kimura, T. Kimura, and K. Ohgushi,  Anisotropic magnetodielectric effect in the honeycomb-type magnet $\alpha-$RuCl$_3$. {\em Phys. Rev. B}. \textbf{95}, 245104 (2017)

\bibitem{jiang}
S. Jiang, J. Shan, and K. Mak,  Electric-field switching of two-dimensional van der Waals magnets. {\em Nat. Mater}. \textbf{17}, 406-410 (2018)

\bibitem{mnps}
H. Chu, C. Roh, J. Island, C. Li, S. Lee, J. Chen, J. Park, A. Young, J. Lee and D. Hsieh,  Linear magnetoelectric phase in ultrathin MnPS$_3$ probed by optical second harmonic generation. {\em Phys. Rev. Lett}. \textbf{124}, 027601 (2020)

\bibitem{hill}
N. Hill, Why are there so few magnetic ferroelectrics$?$ {\em The Journal Of Physical Chemistry B}. \textbf{104}, 6694-6709 (2000)

\bibitem{catalan}
G. Catalan, Magnetocapacitance without magnetoelectric coupling. {\em Applied Physics Letters}. \textbf{88}, 102902 (2006)

\bibitem{dd}
D. Choudhury, P. Mandal, R. Mathieu, A. Hazarika, S. Rajan, A. Sundaresan, U. Waghmare, R. Knut, O. Karis, P. Nordblad and D. Sharma,  Near-room-temperature colossal magnetodielectricity and multiglass properties in partially disordered La$_2$NiMnO$_6$. {\em Phys. Rev. Lett}. \textbf{108}, 127201 (2012)

\bibitem{scott}
W. Eerenstein, N. Mathur, and J. Scott,  Multiferroic and magnetoelectric materials. {\em Nature}. \textbf{442}, 759-765 (2006)

\bibitem{triangularlatt}
S. Shen, J. Wu, J. Song, X. Sun, Y. Yang, Y. Chai, D. Shang, S. Wang, J. Scott and Y. Sun,  Quantum electric-dipole liquid on a triangular lattice. {\em Nat. Commun}. \textbf{7}, 1-6 (2016)

\bibitem{kimura}
T. Kimura, S. Kawamoto, I. Yamada, M. Azuma, M. Takano, and Y. Tokura,  Magnetocapacitance effect in multiferroic BiMnO$_3$. {\em Phys. Rev. B}. \textbf{67}, 180401 (2003)

\bibitem{thinfilm1}
M. Singh, K. Truong, and P. Fournier,  Magnetodielectric effect in double perovskite La$_2$CoMnO$_6$ thin films. {\em Applied Physics Letters}. \textbf{91}, 042504 (2007)

\bibitem{thinfilm2}
J. Lee, L. Fang, E. Vlahos, X. Ke, Y. Jung, L. Kourkoutis, J. Kim, P. Ryan, T. Heeg, M. Roeckerath and Others,  A strong ferroelectric ferromagnet created by means of spin–lattice coupling{\em Nature}. \textbf{466}, 954-958 (2010)

\bibitem{intercalation1}
R. Clement, L. Lomas, and J. Audiere,  Intercalation chemistry of layered iron trithiohypophosphate (FePS$_3$). An approach toward insulating magnets below 90 K. {\em Chemistry Of Materials}. \textbf{2}, 641-643 (1990)

\bibitem{intercalation2}
L. Silipigni, L. Schir\`o, T. Quattrone, V. Grasso, G. Salvato, L. Mons\`u Scolaro, and G. De Luca,  Dielectric spectra of manganese thiophosphate intercalated with sodium ions. {\em J. Appl. Phys}. \textbf{105}, 123703 (2009)







\bibitem{datta}
A. Ghosh, M. Palit, S. Maity, V. Dwij, S. Rana, and S. Datta,  Spin-phonon coupling and magnon scattering in few-layer antiferromagnetic FePS$_3$. {\em Phys. Rev. B}. \textbf{103}, 064431 (2021)



\bibitem{suppli}
See supplementary information for the details of growth, computational methods and other discussions, which includes Refs \cite{micromanip,PhysRevB.47.558,KRESSE199615,Dudarev,DFT-D3,PhysRevB.103.L121108,Ouvard,PhysRevB.73.045112,hippel,jonscher,neupane,wildes,chudnovsky1,chudnovsky2,shih2010quasiparticle,cao2017fully,Lane,Kim1,Takano}.


\bibitem{micromanip}
A. Castellanos-Gomez, M. Buscema, R. Molenaar, V. Singh, L. Janseen, H. Van Der Zant and G. Steele, Deterministic transfer of two-dimensional materials by all-dry viscoelastic stamping. {\em 2D Materials}. \textbf{1}, 011002 (2014)

\bibitem{PhysRevB.47.558}
G. Kresse and J. Hafner, J. {\em Ab initio} molecular dynamics for liquid metals {\em Physical Review B}. \textbf{47}, 558(R) (1993)


\bibitem{KRESSE199615}
G. Kresse and J. Furth M{\"u}ller, Efficiency of ab-initio total energy calculations for metals and semiconductors using a plane-wave basis set. {\em Computational Materials Science}. \textbf{6}, 15 (1996)

\bibitem{Dudarev}
S. L. Dudarev, G. A. Bottom,  S. Y. Savrasov, C. J. Humphreys and A. P. Sutton, Electron-energy-loss spectra and the structural stability of nickel oxide: An LSDA+U study. {\em Phys. Rev. B}. \textbf{57}, 1505 (1998)


\bibitem{DFT-D3}
S. Grimme, J. Antony, S. Ehrlich and H. Krieg, {\em J. Chem. Phys}. \textbf{57}, 154104 (2010)

\bibitem{PhysRevB.103.L121108}
M. Birowska, P. E. Faria Junior, J. Fabian and J. Kuntsmann, Large exciton binding energies in ${\mathrm{MnPS}}_{3}$ as a case study of a van der Waals layered magnet {\em Phys. Rev. B}. \textbf{103}, L121108 (2021)


\bibitem{Ouvard}
G. Ouvrard, R. Brec, and J. Rouxel, {\em Mat.  Res.  Bull.}. \textbf{20}, pp. 1181-1189 (1985)


\bibitem{PhysRevB.73.045112} 
M. Gajdo\v{s}, K. Hummer, G. Kresse, J. Furthm\"uller and F. Bechstedt, Linear optical properties in the projector-augmented wave methodology {\em Phys. Rev. B}. \textbf{73}, 045112 (2006)

\bibitem{hippel}
A. Von Hippel, Dialectrics and Waves. (John Wiley,1954)

\bibitem{jonscher}
A. Jonscher, Dielectric Relaxation in Solids. (Chelsea Dielectrics Press London, 1983)

\bibitem{neupane}
K. Neupane, J. Cohn, H. Terashita and J. Neumeier, Doping dependence of polaron hopping energies in La$_{1- x}$Ca$_x$MnO$_3$ (0 $\le$ x $\le$ 0.15). {\em Physical Review B}. \textbf{74}, 144428 (2006)


\bibitem{wildes}
A. Wildes, D. Lan\c{c}on, M. Chan, F. Weickert, N. Harrison, V. Simonet, M. Zhitomirsky, M. Gvozdikova, T. Ziman and H. R\o{}nnow,  High field magnetization of FePS$_3$. {\em Phys. Rev. B}. \textbf{101}, 024415 (2020)

\bibitem{chudnovsky1}
E. Chudnovsky, D. Garanin, and R. Schilling,  Universal mechanism of spin relaxation in solids. {\em Phys. Rev. B}. \textbf{72}, 094426 (2005)

\bibitem{chudnovsky2}
E. Chudnovsky, and D. Garanin,  Phonon superradiance and phonon laser effect in nanomagnets. {\em Phys. Rev. Lett.}. \textbf{93}, 257205 (2004)

\bibitem{shih2010quasiparticle}
B. Shih, Y. Xue, P. Zhang, M. Cohen and S. Louie, Quasiparticle band gap of ZnO: High accuracy from the conventional G$_0$W$_0$ approach. {\em Phys. Rev. Lett.}. \textbf{105}, 146401 (2010)

\bibitem{cao2017fully}
Cao, H., Yu, Z., Lu, P. \& Wang, L. Fully converged plane-wave-based self-consistent GW calculations of periodic solids. {\em Physical Review B}. \textbf{95}, 035139 (2017)



\bibitem{Lane}
C. Lane, and J.-X. Zhu,  Thickness dependence of electronic structure and optical properties of a correlated van der Waals antiferromagnetic ${\mathrm{NiPS}}_{3}$ thin film. {\em Phys. Rev. B}. \textbf{102}, 075124 (2020)

\bibitem{Kim1}
K. Kim, S. Y. Lim, J.-U. Lee, S. Lee, T. Y. Kim, K. Park, G. S. Jeon, C.-H. Park, J.-G. Park, H. Cheong, Suppression of magnetic ordering in XXZ-type antiferromagnetic monolayer NiPS$_3$. {\em Nature Communications}. \textbf{10}, 345 (2019)

\bibitem{Takano}
Y. Takano, A. Arai, Y. Takahashi, K. Takase, K. Sekizawa,  Magnetic properties and specific heat of new spin glass Mn$_{0.5}$Fe$_{0.5}$PS$_3$ {\em Journal of Applied Physics}. \textbf{93}, 8197-8199 (2003)


\bibitem{kao}
K. Kao, Dielectric phenomena in solids. (Elsevier,2004)
\bibitem{chanda} S. Chanda, S. Saha, A. Dutta, J. Krishna Murthy, A. Venimadhav, S. Shannigrahi, and T. Sinha,  Magnetic ordering and conduction mechanism of different electroactive regions in Lu$_2$NiMnO$_6$. {\em J. Appl. Phys}. \textbf{120}, 134102 (2016)

\bibitem{mott}
N. Mott, and E. Davis,  Electronic processes in non-crystalline materials. (Oxford University Press, 2012)

\bibitem{chen}
X. Chen, C. Zhang, C. Almasan, J. Gardner, and J. Sarrao,  Small-polaron hopping conduction in bilayer manganite La$_{1.2}$Sr$_{1.8}$Mn$_2$O$_7$. {\em Phys. Rev. B}. \textbf{67}, 094426 (2003)

\bibitem{giri}
A. Karmakar, S. Majumdar, and S. Giri,  Polaron relaxation and hopping conductivity in LaMn$_{1-x}$Fe$_x$O$_3$. {\em Phys. Rev. B}. \textbf{79}, 094406 (2009)

\bibitem{rgo}
D. Joung, and S. Khondaker,  Efros-Shklovskii variable-range hopping in reduced graphene oxide sheets of varying carbon sp$^2$ fraction. {\em Phys. Rev. B}. \textbf{86}, 235423 (2012)

\bibitem{holstein}
T. Holstein, Studies of polaron motion: Part II. The \enquote{small} polaron. {\em Annals Of Physics}. \textbf{8}, 343-389 (1959)

\bibitem{seehra}
M. Seehra, and R. Helmick,  Anomalous changes in the dielectric constants of MnF$_2$ near its N\'eel temperature. {\em J. Appl. Phys}. \textbf{55}, 2330-2332 (1984)

\bibitem{infrared}
P. Joy, and S. Vasudevan,  Infrared (700–100 cm$^{-1}$) vibrational spectra of the layered transition metal thiophosphates, MPS$_3$ (M= Mn, Fe and Ni). {\em Journal Of Physics And Chemistry Of Solids}. \textbf{54}, 343-348 (1993)

\bibitem{Sun}
Z. L. Sun \textit{et al.},  Field-induced metal-to-insulator transition and colossal anisotropic magnetoresistance in a nearly Dirac material EuMnSb$_2$ {\em npj Quantum Materials}. \textbf{6}, 94 (2021).

\bibitem{Dgong}
D. Gong, \textit{et al.},  Canted Eu magnetic structure in EuMnSb$_2$ {\em Phys. Rev. B}. \textbf{101}, 224422 (2020).

\bibitem{xzhang}
Xiao-Xiao Zhang \textit{et al.},  Spin Dynamics Slowdown near the Antiferromagnetic Critical Point in Atomically Thin FePS$_3$ {\em Nano Lett.}. \textbf{21}, (12)5045 (2021).

\bibitem{mainak}
D. Vaclavkova, M. Palit, J. Wyzula, S. Ghosh, A. Delhomme, S. Maity, P. Kapuscinski, A. Ghosh, M. Veis, M. Grzeszczyk and Others,  Magnon polarons in the van der Waals antiferromagnet FePS$_3$ {\em Phys. Rev. B}. \textbf{104}, 134437 (2021)

\bibitem{simonson}
J. W. Simonson \textit{et al.},  Magnetic and structural phase diagram of CaMn$_2$Sb$_2$ {\em Phys. Rev. B}. \textbf{86}, 184430 (2012).






\bibitem{park}
Y. Park, K. Song, K. Lee, C. Won, and N. Hur,  Effect of antiferromagnetic order on the dielectric properties of Bi$_2$Fe$4$O$_9$. {\em Applied Physics Letters}. \textbf{96}, 092506 (2010)

\bibitem{schrettle1}
F. Schrettle, S. Krohns, P. Lunkenheimer, J. Hemberger, N. B\"{u}ttgen, H. Von Nidda, A. Prokofiev, and A. Loidl,  Switching the ferroelectric polarization in the S= 1/2 chain cuprate LiCuVO$_4$ by external magnetic fields. {\em Phys. Rev. B}. \textbf{77}, 144101 (2008)

\bibitem{schrettle2}
F. Schrettle, S. Krohns, P. Lunkenheimer, V. Brabers, and A. Loidl,  Relaxor ferroelectricity and the freezing of short-range polar order in magnetite. {\em Phys. Rev. B}. \textbf{83}, 195109 (2011)

\bibitem{shi}
J. Shi, M. Johnson, M. Zhang, P. Gao, and M. Jain,  Antiferromagnetic and dielectric behavior in polycrystalline GdFe$_{0.5}$Cr$_{0.5}$O$_3$ thin film. {\em APL Materials}. \textbf{8}, 031106 (2020)

\bibitem{reviewferroelec}
S. Krohns, and P. Lunkenheimer,  Ferroelectric polarization in multiferroics. {\em Physical Sciences Reviews}. \textbf{4}(9) (2019)

\bibitem{jfscott}Evans, D., Alexe, M., Schilling, A., Kumar, A., Sanchez, D., Ortega, N., Katiyar, R., Scott, J. \& Gregg, J. The nature of magnetoelectric coupling in Pb(Zr,Ti)O$_3$–Pb(Fe,Ta)O$_3$. {\em Advanced Materials}. \textbf{27}, 6068-6073 (2015)




\bibitem{murayama}
C. Murayama, M. Okabe, D. Urushihara, T. Asaka, K. Fukuda, M. Isobe, K. Yamamoto, and Y. Matsushita,  Crystallographic features related to a van der Waals coupling in the layered chalcogenide FePS$_3$. {\em J. Appl. Phys}. \textbf{120}, 142114 (2016)

\bibitem{wildes1}
D. Lan\c{c}on, H. Walker, E. Ressouche, B. Ouladdiaf, K. Rule, G. McIntyre, T. Hicks, H. R\o{}nnow, and A. Wildes,  Magnetic structure and magnon dynamics of the quasi-two-dimensional antiferromagnet FePS$_3$. {\em Phys. Rev. B}. \textbf{94}, 214407 (2016)

\bibitem{budniak}
A. K. Budniak, S. J. Zelewski, M. Birowska, T. Wo\'{z}niak, T. Bendikov, Y. Kauffmann, Y. Amouyal, R. Kudrawiec and E. Lifshitz,  Spectroscopy and Structural Investigation of Iron Phosphorus Trisulfide—FePS$_3$. {\em Advanced Optical Materials}. \textbf{10}, 2102489 (2022)

\bibitem{jernberg}
P. Jernberg, S. Bjarman, and R. W\"appling,  FePS$_3$: A first-order phase transition in a “2D” Ising antiferromagnet. {\em Journal Of Magnetism And Magnetic Materials}. \textbf{46}, 178-190 (1984)

\bibitem{nauman}
M. Nauman, D. Kiem, S. Lee, S. Son, J. Park, W. Kang, M. Han, and Y. Jo,  Complete mapping of magnetic anisotropy for prototype Ising van der Waals FePS$_3$. {\em 2D Materials}. \textbf{8}, 035011 (2021)

\bibitem{domainising}
C. Sarkis, S. S\"{a}ubert, V. Williams, E. Choi, T. Reeder, H. Nair, and K. Ross,  Low-temperature domain-wall freezing and nonequilibrium dynamics in the transverse-field Ising model material CoNb$_2$O$_6$. {\em Phys. Rev. B}. \textbf{104}, 214424 (2021)



\bibitem{wahab}
D. Abdul-Wahab, E. Iacocca, R. Evans, A. Bedoya-Pinto, S. Parkin, K. Novoselov, and E. Santos,  Domain wall dynamics in two-dimensional van der Waals ferromagnets. {\em Applied Physics Reviews}. \textbf{8}, 041411 (2021)

\bibitem{mnps3domain}
I. Alliati, R. Evans, K. Novoselov, and E. Santos,  Relativistic domain-wall dynamics in van der Waals antiferromagnet MnPS$_3$. {\em Npj Computational Materials}. \textbf{8}, 1-9 (2022)



\bibitem{Ellenore}
E. Geraffy, S. Zuri, M. M. Rybak, F. Horani, A. K. Budniak, Y. Amouyal, M. Birowska, E. Lifshitz, Crystal anisotropy  implications on the intrinsic magnetic and optical properties in van der Waals FePS$_3$. {\em https://arxiv.org/pdf/2208.10890.pdf}

\bibitem{amirabbasi}Amirabbasi, M. \& Kratzer, P. Orbital and magnetic ordering in single-layer FePS$_3$: A DFT+ U study. {\em Physical Review B}. \textbf{107}, 024401 (2023)




\bibitem{magda1}
C. Autieri, G. Cuono, C. Noce, M. Rybak, K. Kotur, C. Agrapidis, K. Wohlfeld, and M. Birowska,  Limited Ferromagnetic Interactions in Monolayers of MPS$_3$ (M= Mn and Ni). {\em The Journal Of Physical Chemistry C}. \textbf{126}, 6791-6802 (2022)






 










\end{thebibliography}

\newpage 

\end {document}


\title{Supplementary Information for: Anisotropic Magnetodielectric Coupling in Layered Antiferromagnetic FePS$_3$}


\author{Anudeepa Ghosh}
\affiliation{School of Physical Sciences, Indian Association for the Cultivation of Science, 2A \& B
Raja S. C. Mullick Road, Jadavpur, Kolkata - 700032, India
}
\author{Magdalena Birowska}
\affiliation{Faculty of Physics, Institute of Theoretical Physics, University of Warsaw, Pasteura 5, 02093, Warsaw, Poland}

\author{Pradeepta Kumar Ghose}
\affiliation{School of Physical Sciences, Indian Association for the Cultivation of Science, 2A \& B
Raja S. C. Mullick Road, Jadavpur, Kolkata - 700032, India
}

\author{Miłosz Rybak}
\affiliation{Department of Semiconductor Materials Engineering, Faculty of Fundamental Problems of Technology, Wrocław University of Science and Technology, Wybrzeże Wyspiańskiego 27, PL-50370 Wrocław, Poland}

\author{Sujan Maity}
\affiliation{School of Physical Sciences, Indian Association for the Cultivation of Science, 2A \& B
Raja S. C. Mullick Road, Jadavpur, Kolkata - 700032, India
}

\author{Somsubhra Ghosh}
\affiliation{School of Physical Sciences, Indian Association for the Cultivation of Science, 2A \& B
Raja S. C. Mullick Road, Jadavpur, Kolkata - 700032, India
}

\author{Bikash Das}
\affiliation{School of Physical Sciences, Indian Association for the Cultivation of Science, 2A \& B
Raja S. C. Mullick Road, Jadavpur, Kolkata - 700032, India
}

\author{Koushik Dey}
\affiliation{School of Physical Sciences, Indian Association for the Cultivation of Science, 2A \& B
Raja S. C. Mullick Road, Jadavpur, Kolkata - 700032, India
}

\author{Satyabrata Bera}
\affiliation{School of Physical Sciences, Indian Association for the Cultivation of Science, 2A \& B
Raja S. C. Mullick Road, Jadavpur, Kolkata - 700032, India
}


\author{Suresh Bhardwaj}
\affiliation{UGC-DAE Consortium for Scientific Research, University Campus, Khandwa Road, Indore-452001,India}

\author{Shibabrata Nandi}
\affiliation{Forschungszentrum J\"{u}lich GmbH, J\"{u}lich Centre for Neutron Science (JCNS-2) and Peter Gr\"{u}nberg Institut (PGI-4), JARA-FIT, 52425 J\"{u}lich, Germany}
\affiliation{RWTH Aachen, Lehrstuhl f\"{u}r Experimentalphysik IVc, J\"{u}lich-Aachen Research Alliance (JARA-FIT), 52074 Aachen, Germany}

\author{Subhadeep Datta}
\affiliation{School of Physical Sciences, Indian Association for the Cultivation of Science, 2A \& B
Raja S. C. Mullick Road, Jadavpur, Kolkata - 700032, India
}

\maketitle
\setcounter{tocdepth}{3} 
\tableofcontents
\newpage

\section{Sample preparation and measurement}

\begin{figure}[h!]
\centerline{\includegraphics[scale=0.6, clip]{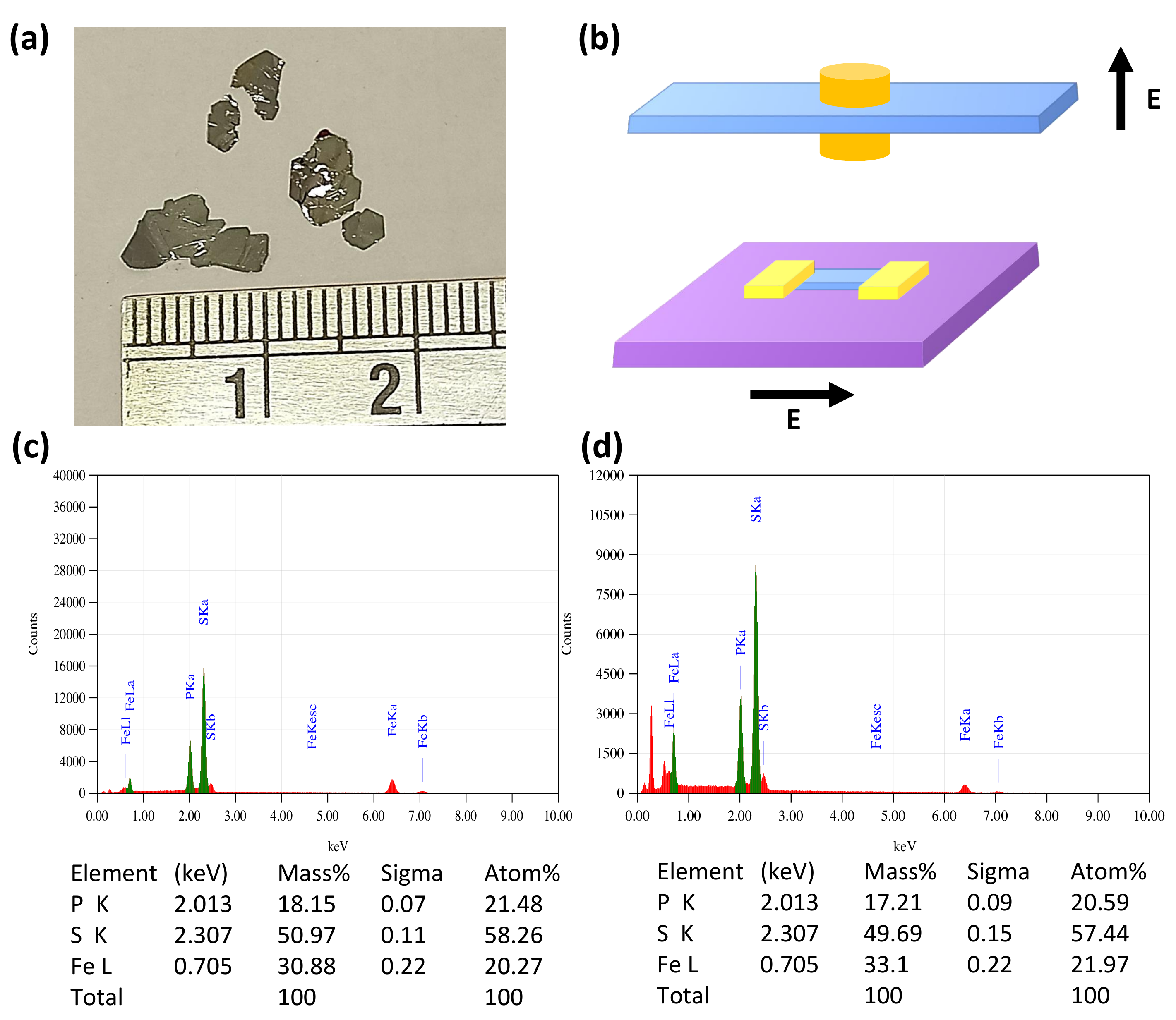}}
\caption{Iron phosphorus trisulfide (FePS$_3$): (a) Crystals grown \textit{via} CVT. Biggest crystals reached size of $\thicksim$ 1 cm. (b) Schematics showing in-plane and out-of-plane measurement schemes of dielectric constant. (c)-(d) Energy dispersive x-ray spectroscopy (EDS) measurementis performed in a JEOL JSM-7500F scanning electron microscope (SEM) to confirm the composition and the homogeneity of the single crystals grown \textit{via} (c) CVT and (d) box furnace methods. 
\label{measurement}}
\end{figure}

FePS$_{3}$ was grown in two ways. Pure Fe, P, S (Alpha Aesar) powder in stoichiometric amounts with iodine as transport agent were sealed in two different quartz tube under vacuum ($\approx$ 10$^{-6}$ mbar). One of the tubes was placed in a single-zone box furnace at a temperature of 750$^{0}$C for 7 days and then cooled at a rate of 1$^{0}$C per minute. Small crystals of size about 0.5 mm was formed. The other quartz tube was placed in a two-zone furnace and growth was done \textit{via} chemical vapor transport (CVT) method. The hot-zone temperature was set to 750$^{0}$C and cold-zone temperature to 650$^{0}$C and kept for 8 days. Larger sized crystals of dimensions up to 10 mm were formed Fig. \ref{measurement}(a). Both the crystals were characterized \textit{via} two types of X-ray diffraction (XRD)- powder XRD (Rigaku Smart Lab x-ray diffractometer with Cu-K$\alpha$ radiation) and single crystal XRD (Bruker Smart Apex 2 CCD diffractometer) and Raman Spectroscopy (Horiba T64000). Energy dispersive x-ray spectroscopy (EDS) measurement was performed in a JEOL JSM-6010LA scanning electron microscope (SEM) to confirm the composition and the homogeneity of the single crystals as well as bulk and few-layer flakes transferred onto SiO$_2$/Si wafer [Fig. \ref{measurement} (c)-(d)]. Magnetic susceptibility (DynaCool, Quantum Design) measurements on bulk samples (DC magnetic field 0.1 T, 0.5 T and 1 T, AC rms field 7 Oe, frequencies ranging from 11 to 9999 Hz) were done to confirm the antiferromagnetic nature (T$_{N}$ = 118 K).

Temperature dependent (4 K - 300 K) dielectric spectroscopy (LCR meter, model: Agilent E4980A) with varying frequency from 10 Hz to 1 MHz and magnetic field (B) were performed following the \enquote{parallel-plate} geometry (Au electrodes) for the in-plane and out-of-plane measurement on the transferred bulk material. The out-of-plane measurement was done on a crystal with top and bottom connections made on a bulk transferred flake \textit{via} optical and electron beam lithography (Au electrodes). For in plane measurement, FePS$_{3}$ was exfoliated $\textit{via}$ standard scotch tape method and a bulk flake was stamped on top of predefined two-probe gold electrodes \textit{via} micromanipulation technique \cite{micromanip}. A schematic of the two measurement geometries are shown in Fig. \ref{measurement}(b). All units, except otherwise specified, are either arbitrary or normalised since the samples lacked specific shape. 

Characteristic spin-phonon coupled modes were tracked with lowering temperature \textit{via} Raman scattering (see \cite{datta}).


\section{Computational Details}

\begin{figure}[h!]
\centerline{\includegraphics[scale=0.4, clip]{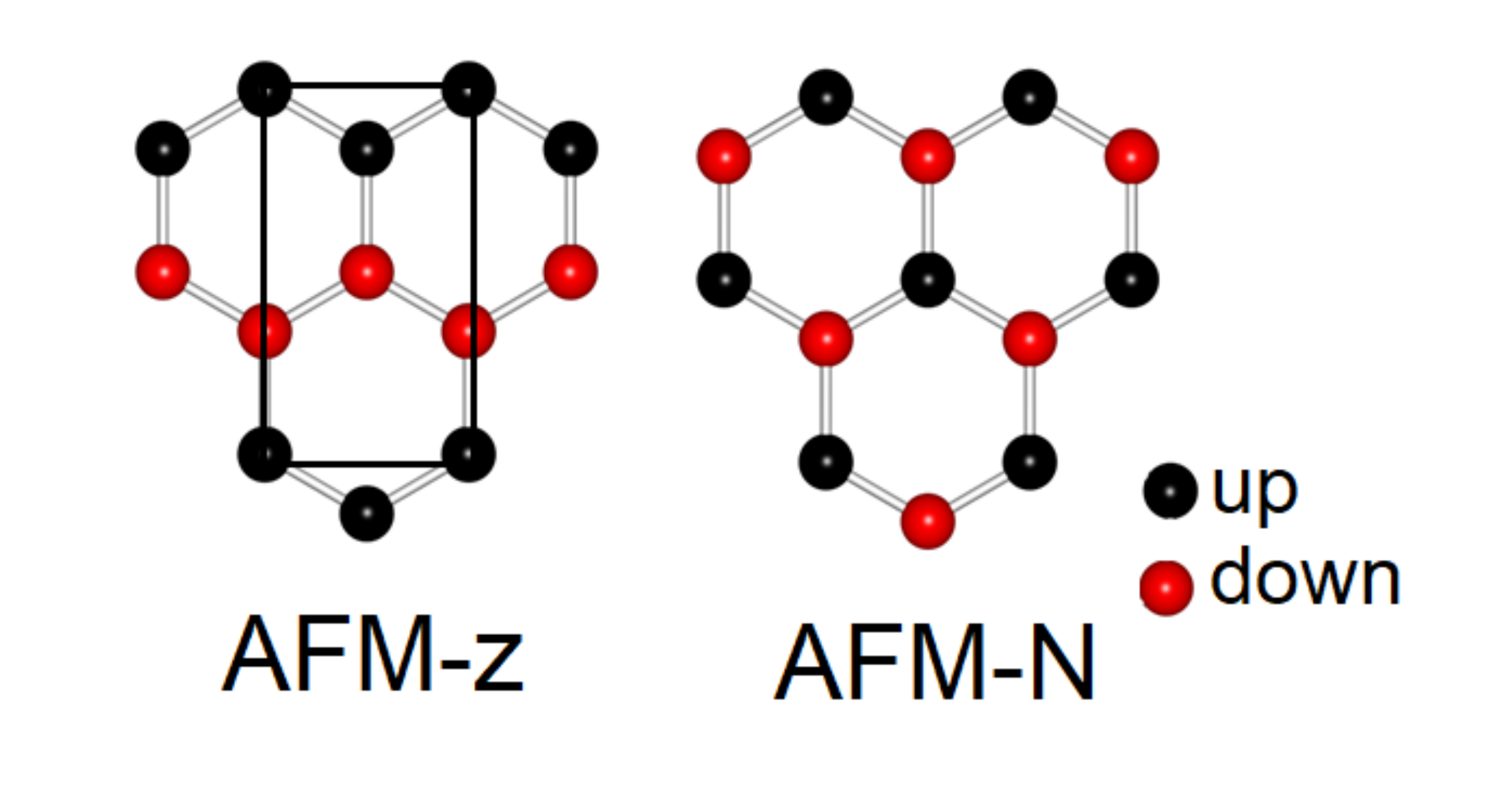}}
\caption{AFM-z and AFM-N magnetic orderings are presented. The black and red balls represent the spin up and down directions of the Fe atoms, respectively. The black solid line denotes the planar rectangular cell used in the calculations.
\label{structure}}
\end{figure}

The calculations were performed in the framework of spin-polarized DFT, using the  projector-augmented-wave (PAW) based Vienna \textit{ab initio} Simulation Package (VASP) \cite{PhysRevB.47.558,KRESSE199615}. DFT+U formalism  proposed by Dudarev \cite{Dudarev} was adopted to properly characterize on-site Coulomb repulsion between $3d$ electrons of Fe ions, by using effective Hubbard U (U$_{eff}$ =U-J, where J=1eV). We adopted two benchmark values equal to U=2.6 eV and U=5.3 eV as previously discussed in paper \cite{Budniak}. The dispersive forces were included within the semi-empirical Grimme approach (DFT-D3) \cite{DFT-D3}. FePS$_3$ bulk material exhibits the AFM zigzag (AFM-z) order within the layer, and the adjacent layers are antiferromagnetically aligned.  Note that, the magnetic supercell of bulk FePS$_3$ is not commensurate with its primitive  lattice supercell (see paper \cite{Budniak} and discussion therein). In order to make the dielectric properties computationally feasible, we have chosen the smaller supercell for the bulk systems which includes the same magnetic state in respect to the adjacent layers and contains 20 atoms (see black rectangular in Fig. \ref{structure}), similarly as reported in  \cite{PhysRevB.103.L121108}. 

The lattice parameters have been fixed to experimental lattice parameters equal to a=5.947 $\AA{}$, b=10.3 $\AA{}$, c= 6.722 $\AA{}$ \cite{Ouvard}. A cutoff of 400 eV was chosen for the plane-wave basis set. A k-mesh of $10\times6\times9$ was taken to sample the first Brillouin zone on $\Gamma$-centered symmetry reduced Monkhorst-Pack mesh. The position of atoms were relaxed until the maximal force per atom is less than $10^{-3}$ eV/$\AA{}$. 

The static dielectric properties were calculated by means of density functional perturbation theory implemented in the VASP. The dielectric constant represents the macroscopic static response containing the electronic $\varepsilon_{\infty }$ and ionic $\varepsilon_{ion}$ contributions. The electronic contribution $\varepsilon_{\infty }$ (macroscopic optical dielectric constant) was calculated  in the Independent Particle (IP) approach  including the local field effects  \cite{PhysRevB.73.045112} on the top of PBE+U approach. 
 Although, the IP method is known to  overestimate the electronic contribution of  dielectric constant of semiconducting bulk crystals by up to 20 $\%$ \cite{PhysRevB.73.045112},  the post-DFT approaches such as single-shot G$_0$W$_0$ approximation are still restricted to few-atoms systems and require thousands of the conduction bands to reach convergence \cite{shih2010quasiparticle,cao2017fully}.  In the case of the ionic contribution (phononic part at low frequency regime), the relaxation of the electronic degrees of freedom stops when the total energy change was set to $10^{-8}$ eV, and the norms of all the forces were set to be smaller than $10^{-9}$ eV/$\AA{}$. The force-constant matrices and internal strain tensors were calculated within the finite deference method (FDM). The above mentioned thresholds were important to ensure that the three imaginary phonon modes have negligible values, and do not impact on the ionic part to the dielectric constants.

\section{Results and discussion}

\subsection{Dielectric Spectroscopy}

\subsubsection{Out-of-plane spectra:} 
\begin{figure}[ht]
\centerline{\includegraphics[scale=0.6, clip]{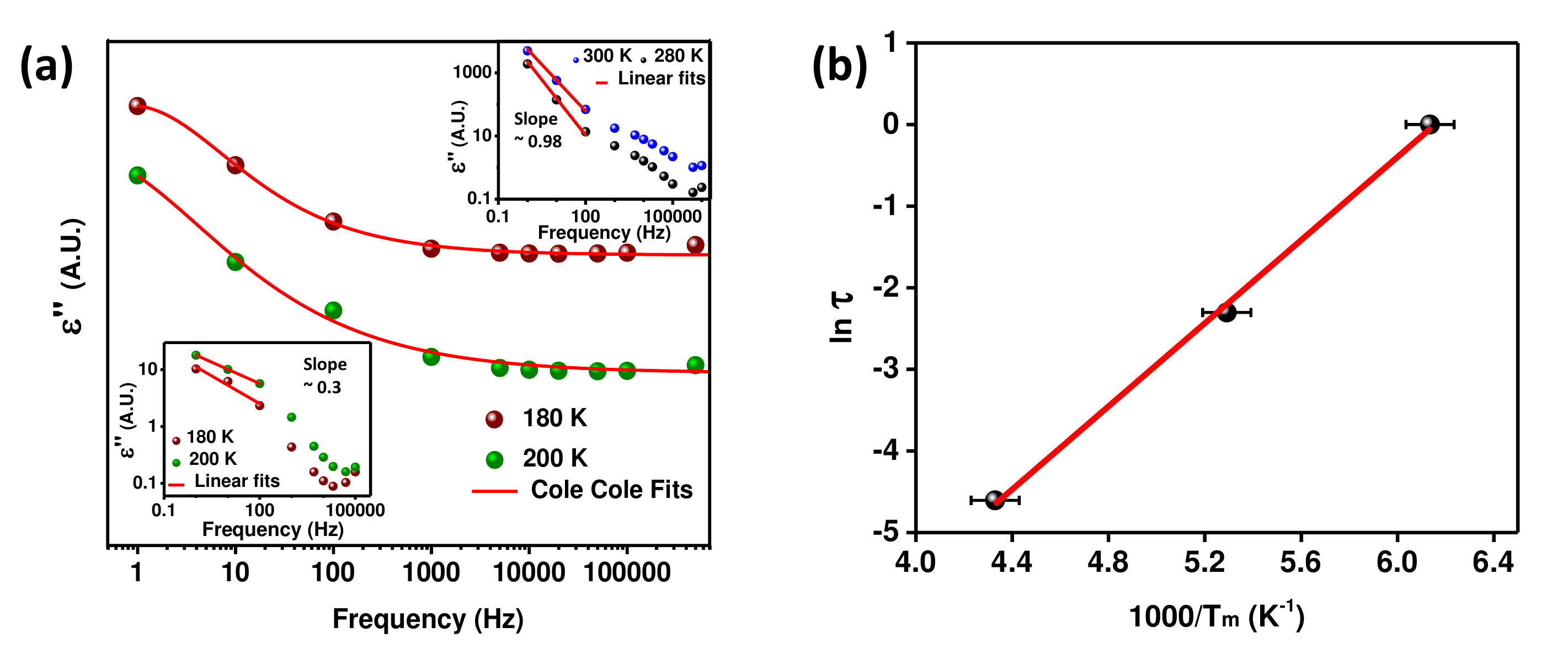}}
\caption{Dielectric spectroscopy in the antiferromagnetic transition region for out-of-plane geometry. (a) Cole Cole fit in region A for two temperatures. Top inset shows MW relaxations for higher temperatures(region B) marked by slope $\thicksim$ -1 in log $\varepsilon''$ vs log frequency plots. Lower inset shows  similar plot where MW relaxations are not seen and slope $\thicksim$ -0.3 (b) Relaxation time as calculated from maximum in tan $\delta$.
\label{dielectric}}
\end{figure}

The Maxwell-Wagner relaxations can be identified by its characteristic $f^{-1}$ dependence of the imaginary part of dielectric constant data ($\varepsilon''$) at lower frequencies \cite{hippel}. As seen in the upper inset of Fig. \ref{dielectric}(a), the slope of $\rm{log}(\varepsilon''$) vs $\rm{log}(f)$ is found to be (-1) in region marked as B in main text [see Fig. 2(a)] clearly suggesting the presence of the Maxwell-Wagner (MW) relaxation. 

Even though MW relaxation model fit well in the higher temperature-lower frequency regime, it fails to fit the data in region A in main text [see Fig. 2(a)] as shown in the lower inset of Fig. \ref{dielectric}(a). To understand the origin of higher frequency data in the interim temperature regime (\textgreater 150 K) the data was fitted with Debye-like model of the form:

\begin{equation}
\varepsilon (\omega) = \varepsilon_{\infty} + \frac{\varepsilon_{s} - \varepsilon_{\infty} }{1+(i\omega\tau)^{1-\alpha}} 
\end{equation}
					
This is the Cole-Cole equation and it fits the data well in region A as shown in Fig. \ref{dielectric}(a)]. This explains the experimental data in the entire temperature and frequency range probed.

For Debye-like relaxation, the relaxation time and activation energy can be determined from the relation \cite{jonscher,neupane}:

\begin{equation}
\tau = \tau_0 \exp(E/k_BT)
\end{equation}

where symbols have their usual meaning. Values of E and $\tau_0$ were determined from the slopes and intercepts, respectively, of linear fits of semilog plot of $\omega$ vs 1000/$T_{max}$ where $T_{max}$ corresponds to the temperature in tan $\delta$ with $\omega = 1/ \tau$ being the corresponding frequency [Fig. \ref{dielectric}(b)]. The values of E and $\tau_0$ were found to be 219 meV and 1.5 x $10^{-7}$ s respectively.  \\

 \

\subsubsection{Magnetodielectric coupling:}

\begin{figure}[h!]
\centerline{\includegraphics[scale=0.6, clip]{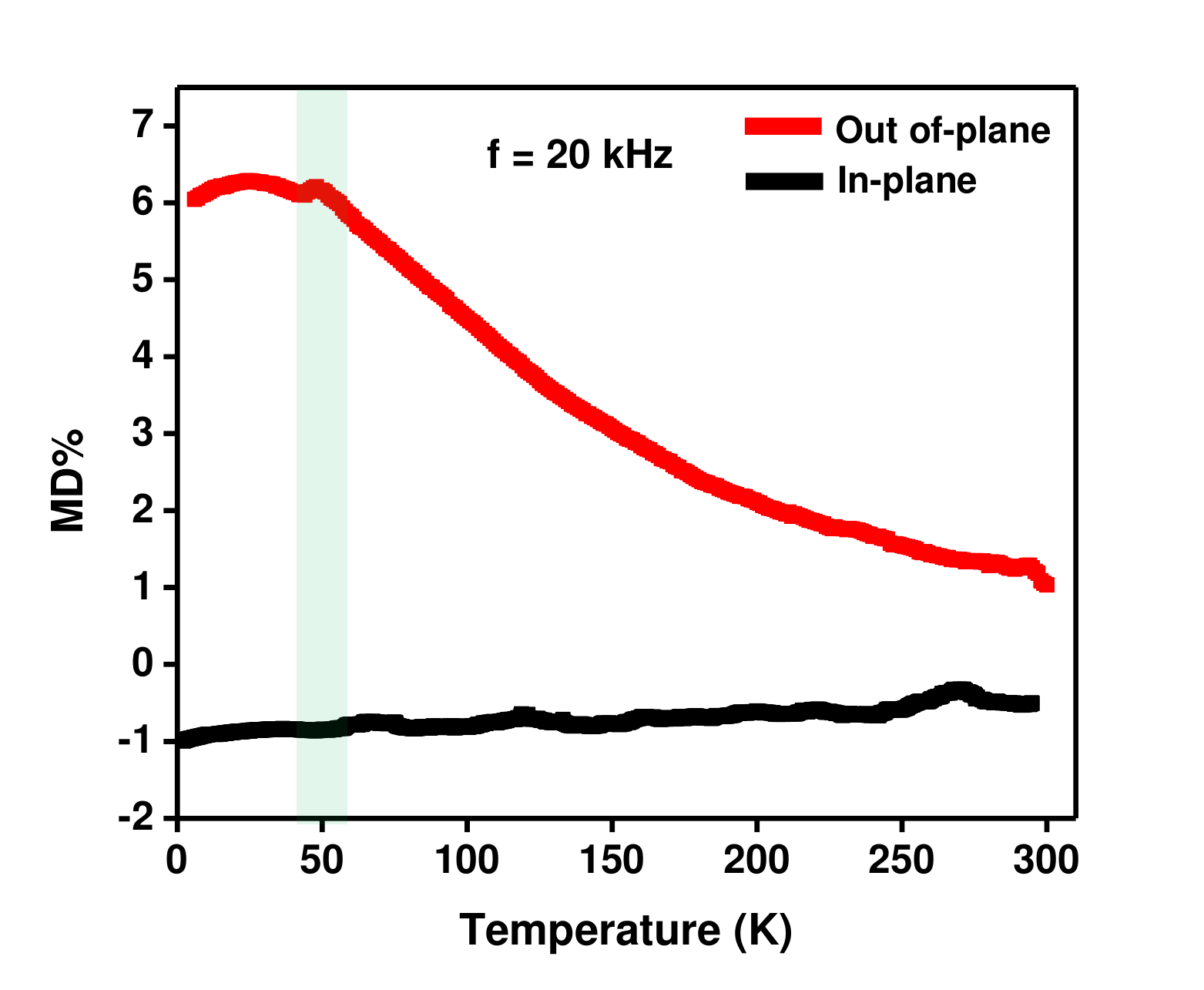}}
\caption{Magnetodielectric coupling measured in two device geometries for the frequency 20 kHz. The out-of-plane curve was subjected to a field of 5 T while the in-plane one was measured at 0.1 T. The out-of-plane curve shows a peak around 50 K.
\label{mdc}}
\end{figure}

We have demonstrated the magnetodielectric coupling (MDC) with temperature in FePS$_3$ by applying a constant magnetic field (\textbf{H}) parallel to \textbf{E} (\textbf{H} $\mid$ \textbf{E}) for both, \enquote{in-plane} and \enquote{out-of-plane} configuration (Fig. \ref{mdc}). (Note that the direction of applied magnetic field is different from the sample in main text.) For the \enquote{out-of-plane} geometry, magnetodielectric effect (MD = ($\varepsilon'(H)$ - $\varepsilon'(0)$)/$\varepsilon'(0)$)) with a maximum of 6\% at 4 K is observed for H = 5 T. While below $T_{\mathrm{N}}$, MD shows a slight temperature dependent variation ($\sim$ 0.5\%), it decreases sharply with increasing temperature above $T_{\mathrm{N}}$. For higher temperature regime, MW effects can contribute to the observed positive MDC \cite{catalan}, which however is not present below $T_{\mathrm{N}}$ (AFM phase). On the other hand, the MDC in the \enquote{in-plane} direction is (-1\%) at lowest temperature (T = 4 K) and remains relatively constant with temperature. Taking cue from our previous works \cite{datta}, we believe that an intrinsic coupling between the spin and lattice below $T_{\mathrm{N}}$ may modify the dielectric properties at the phase transition and is the leading contributor to MDC in AFM phase. Note that magnetostriction effects, which usually sets in at very high fields ($\geq$ 25T ) for FePS$_3$ \cite{wildes}, can be ruled out as a possibility in this case.

\subsubsection{Spin-phonon coupling:} To explain the anomaly in spin-phonon coupling around 40 K [Fig. 1(b) (see main text)] we resort to the theory of spin-phonon dynamics has been developed by Chudnovsky \textit{et al.} \cite{chudnovsky1,chudnovsky2} where the Hamiltonian \^{H}$_{s-ph}$ is given by:

\begin{equation}
\hat{H}_{s-ph} = -\hbar S \cdot \Omega
\text{,}\quad
\Omega= \frac{1}{2} \triangledown \times \dot{u}(r)   
\end{equation}

where S is the spin of the atom, $\Omega$ is the angular velocity of the local rotation of the crystal, and u is the phonon displacement field. The crystal field, governed by the magnetic anisotropy, that determines the spin states of an atom in a solid, is defined in a local coordinate frame coupled to the crystal axes. This can be used to understand relaxation of a spin cluster due to local rotations of differently frozen domains. The magnitude of spin-phonon coupling ($\propto \Delta\omega$) is governed by local magnetic anisotropy which changes locally due to rotation within the domains which changes $\Omega$ and when the domain freezes around 40 K it result in a change in the strength of spin-phonon coupling as seen in Fig. 1(b) of main text.
 

\subsection{AC Susceptibility Measurements}

\begin{figure}[h!]
\centerline{\includegraphics[scale=0.5, clip]{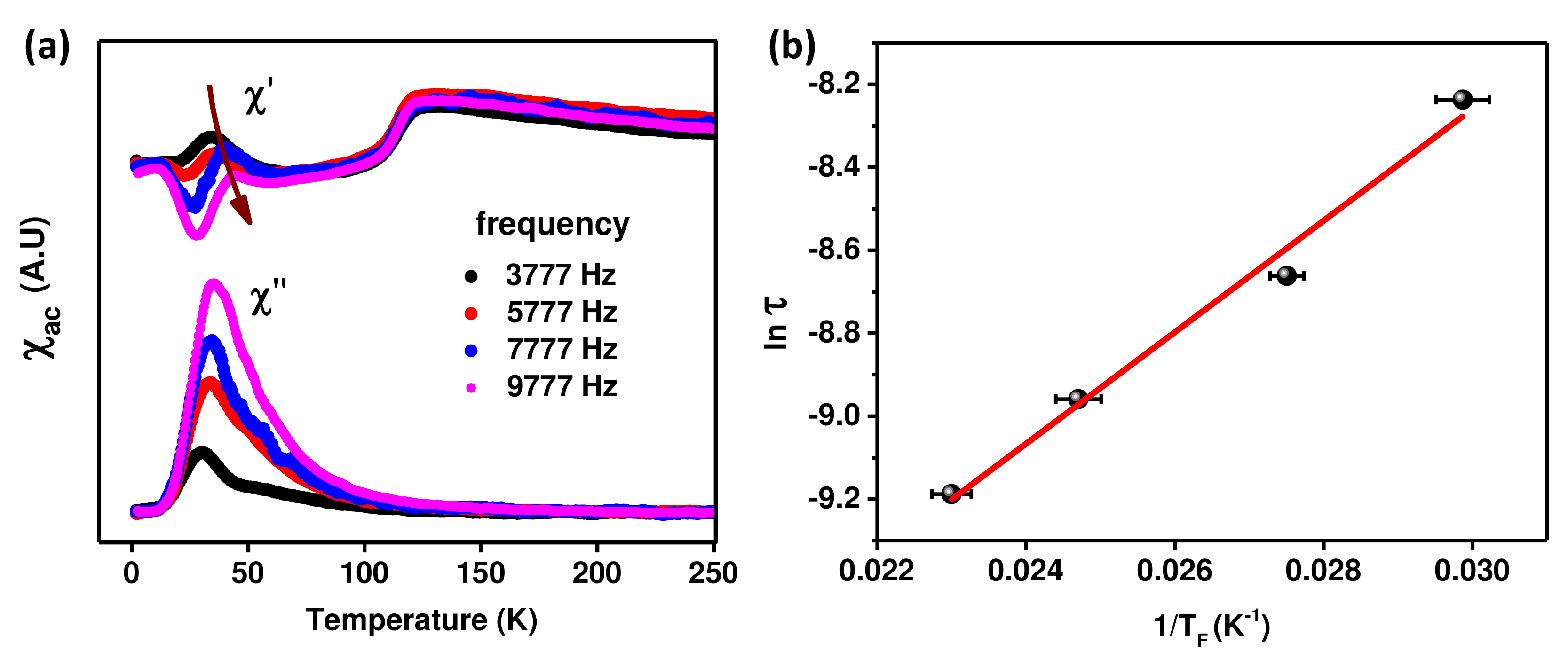}}
\caption{(a) AC susceptibility measurements $\chi_{ac}$ at different frequencies. The real ($\chi'$) and imaginary ($\chi''$) susceptibility are shifted arbitrarily along $\chi_{ac}$ axis. Prominent peaks in $\chi'$ evolve around 30 K and show large shift towards higher temperature with increasing frequency as marked by arrow. Another set of peaks are seen just below the ones marked. $\chi''$ shows a single prominent set of peaks with large temperature shift with increasing frequency. (b) Arrhenius fit to the set of ac peaks $\chi'$ marked by arrow in (a).
\label{ac}}
\end{figure}

Using the Arrhenius equation defined as:

\begin{equation}
\tau = \tau_0 \exp(E/k_BT)
\end{equation}

for peaks marked by arrow in Fig. 3(d) in main text, the values of E and $\tau_0$ were found to be 11.6 meV and 4.6 x $10^{-6}$ s respectively [Fig. \ref{ac}].


\subsection{Additional theoretical considerations}
\subsubsection{The static dielectric constants as a function of the lattice parameters’ change}
\begin{figure}[h!]
\centerline{\includegraphics[scale=0.6, clip]{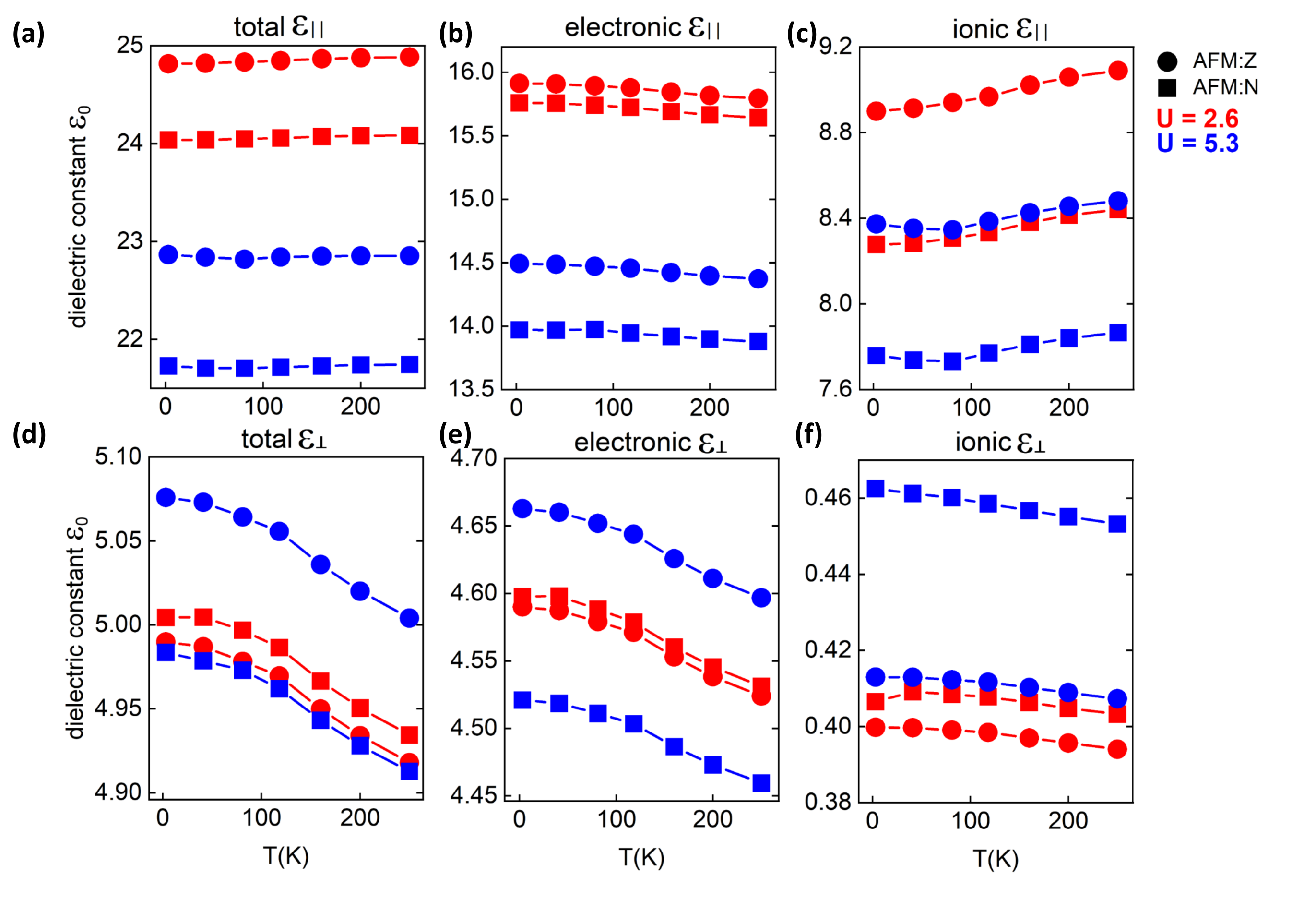}}
\caption{ Dielectric constant as a function of the lattice parameters' changes at various temperatures. (a-c) The in-plane $\varepsilon_{\parallel}$ and (d-f) out of-plane $\varepsilon_{\perp}$ contributions of static dielectric constant for two different magnetic phases and Hubbard U parameters. The electronic and ionic contributions are presented  on the middle and right side of the graphs, respectively. The in-plane component $\varepsilon_{\parallel}$ is the average of the x and y contributions of the static dielectric constant within the ab plane of the layer, where the  is a out-of plane contribution (parallel to c direction).
\label{dielectric1}}
\end{figure}

\begin{figure}[h!]
\centerline{\includegraphics[scale=0.56, clip]{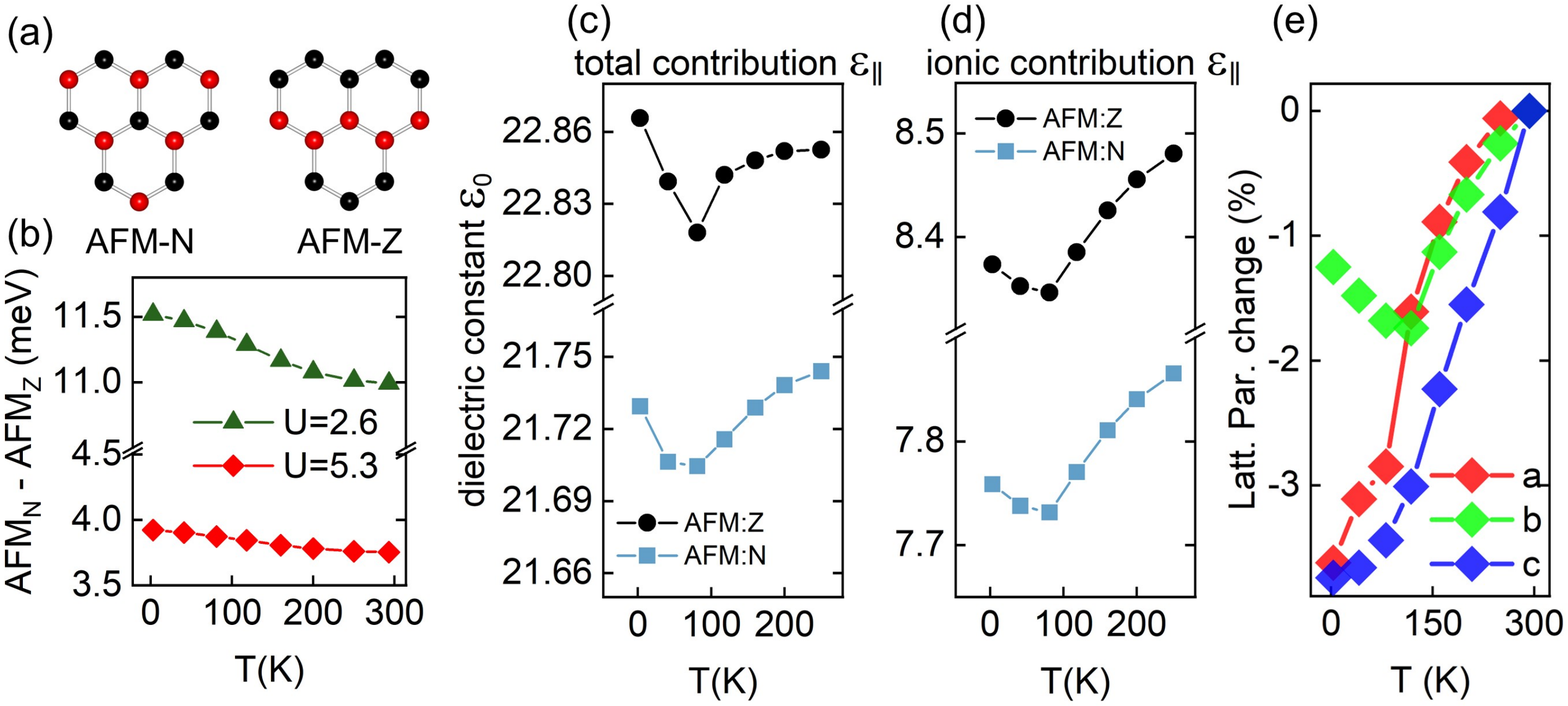}}
\caption{ (a) AFM-z and AFM-N magnetic orderings are presented. The black and red balls represent the spin up and down directions of the Fe atoms, respectively. (b) The energy difference between the magnetic AFM-N and AFM-z phases as a function of the temperature, which reflects the changes in lattice parameters, however only for U = 5.3 eV, whereas for U = 2.6 eV the kink is not observed. (c) The temperature dependent in-plane dielectric constant with its (d) ionic contribution (for U = 5.3 eV) are plotted as the function of the lattice parameters’ changes presented in (e). Relative deviations of the lattice parameters with respect to the room temperature values are taken from the XRD measurements \cite{Murayama}
\label{oldtheory}}
\end{figure}

Here, we present in details the results considering the energy difference between the two lowest magnetic phases: AFM-z and AFM-N (see Fig. \ref{dielectric1}(a)) as a function of the lattice parameters extracted from experimental measurements at elevated temperatures (model I). Regarding, the dielectric properties,  the in-plane contributions depend on the magnetic order, and stronger differences are obtained for larger value of Hubbard U parameter, whereas the out-of plane contributions are not sensitive to magnetic order (see Fig. \ref{oldtheory}). In particular, the AFM-N type of order exhibits lower values of the in-plane dielectric constant by about 5\% for U = 5.3 eV (3\% for U = 2.6 eV) in respect to the AFM-z phase, whereas the corresponding difference for the out-of plane contributions are around tenths of the dielectric constant and can be considered as negligible (see Table \ref{tab:dielectric}).

\begin{table}[h!]
\caption{In-plane ($\varepsilon_{\parallel}$) and out-plane ($\varepsilon_{\perp}$) contributions of dielectric constant $\varepsilon_0$ computed for AFM-z and AFM-N magnetic phases of bulk systems. The electronic ($\varepsilon_\infty$) and ionic ($\varepsilon_{ion}$) contributions are presented.}
\label{tab:dielectric}
\begin{ruledtabular}
\begin{tabular}{cc|cc|cc|cc}
U [eV]&& \multicolumn{2}{c}{ $\varepsilon_0$  }  & \multicolumn{2}{c}{ $\varepsilon_\infty$ } & \multicolumn{2}{c}{ $\varepsilon_{ion}$  }\tabularnewline
& & ${\parallel}$ & ${\perp}$ & ${\parallel}$ & ${\perp}$& ${\parallel}$ & ${\perp}$\tabularnewline
 \hline
U=2.6 &AFM-z  &24.9 & 4.9 & 15.8& 4.5 & 9.1 & 0.4  \tabularnewline
 &AFM-N &24.1 &4.9 &15.6  & 4.5 & 8.4 &0.4 \tabularnewline
\hline
 U=5.3 &AFM-z&22.9 & 5.0 & 14.4& 4.6 & 8.5 & 0.4  \tabularnewline
&AFM-N &21.7 &4.9 &13.9  & 4.5 & 7.9 &0.5 \tabularnewline
\end{tabular}
\end{ruledtabular}
\end{table}

The energy difference between the magnetic phases is nearly independent on lattice parameters (see Fig. \ref{oldtheory}(b)). This energy difference indicates that the thermal energy could flip the magnetic order from AFM-z to AFM-N, and these two phases could coexist  at some higher temperature. However, it is difficult to evaluate this critical temperature within DFT+U studies, as the energy difference for FePS$_3$ system depends strongly on U parameter. In particular, the AFM-N\'{e}el phase is greater by 3.9 meV for U = 5.3 eV and around 11 meV for U = 2.6 eV in respect to AFM-z phase, which corresponds to temperatures of equal to 45 K and 133 K, respectively.   Interestingly, these two magnetic phases were considered to be energetically degenerated for monolayer of NiPS$_3$ \cite{Lane}. This is consistent with the suppression of long-range order observed by Raman spectroscopy due to strong magnetic fluctuations for this system \cite{Kim1}. Note that, disorder state with competing magnetic interactions have been also observed for mixed compound Mn$_{1-x}$Fe$_x$S$_3$ \cite{Takano}.

\subsubsection{Description of the model III}

\begin{figure}[ht]
\centerline{\includegraphics[scale=0.8, clip]{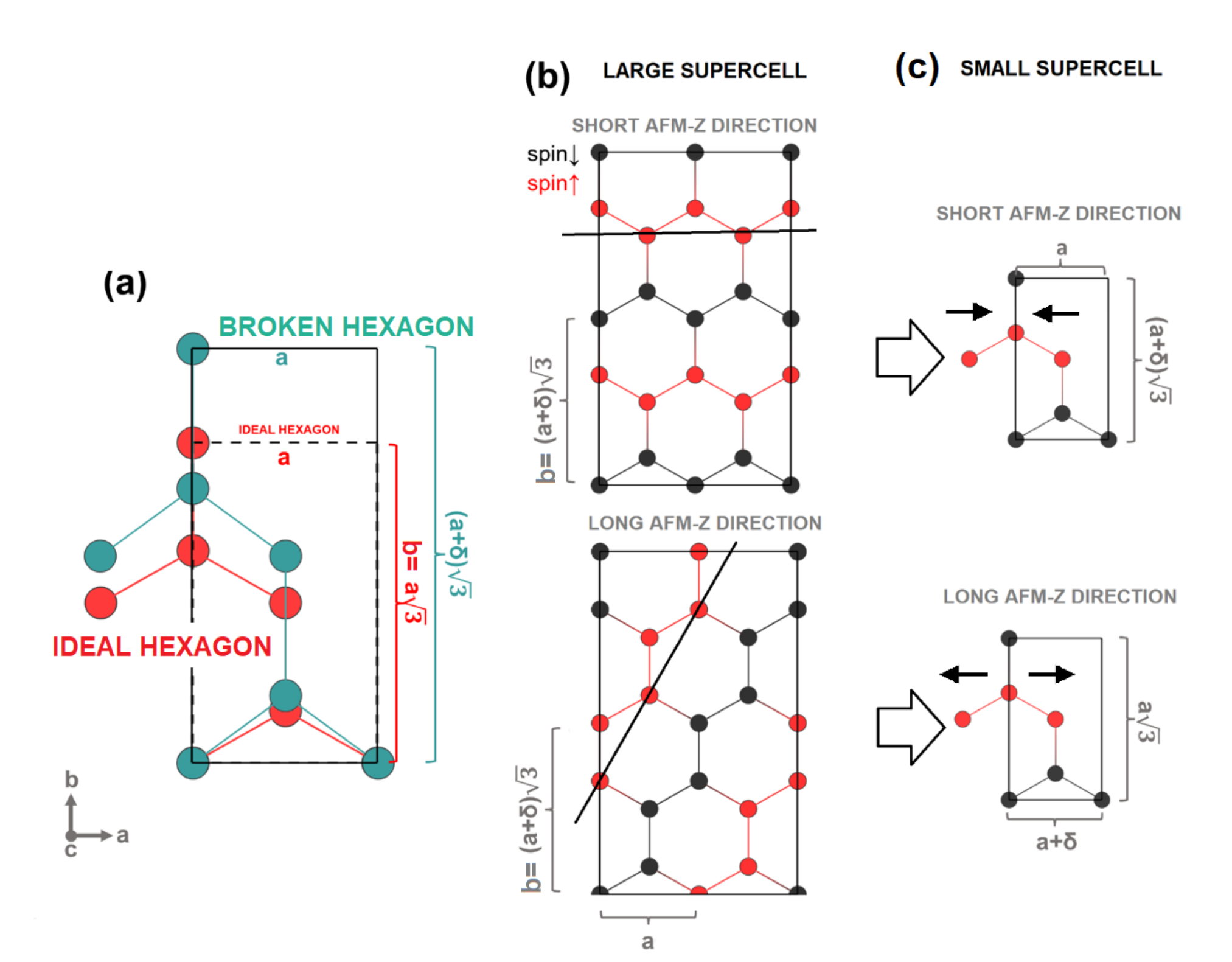}}
\caption{An oversimplified model representing the change in orientation of AFM-z within the layer.
\label{theorymodel}}
\end{figure}
To capture the two different alignment of the AFM-z ordering within the supercell approach at least planar 2$\times$2 supercell of primitive hexagonal one need to be employed. Such supercell contains  40 atoms, and calculation regarding the ionic contribution of the dielectric constant are very demanding. Hence, we employ oversimplified model, that capture the change of the orientation of AFM-z phase within the  smaller rectangular supercell, containing 20 atoms as presented in Fig. \ref{theorymodel}. Note, that using the smaller supercell there exist only one possible alignment of the AFM-z.  As it was reported recently, the in-plane structural anisotropy is observed for the FePS$_3$ structure, thus two AFM-z alignments on the Fig. \ref{theorymodel}(b) are not equivalent anymore, because the hexagonal symmetry of the lattice is broken ($b \neq \sqrt{3}*a$) \cite{Ellenore,amirabbasi} . Therefore, by breaking the hexagonal symmetry within the layer (using parameter $\delta$ = 0.2\%) we can map the change of the directions of AFM-z phase.

\section{Magnetic field induced quantum fluctuation at low temperature}

\begin{figure}[h!]
\centerline{\includegraphics[scale=0.38, clip]{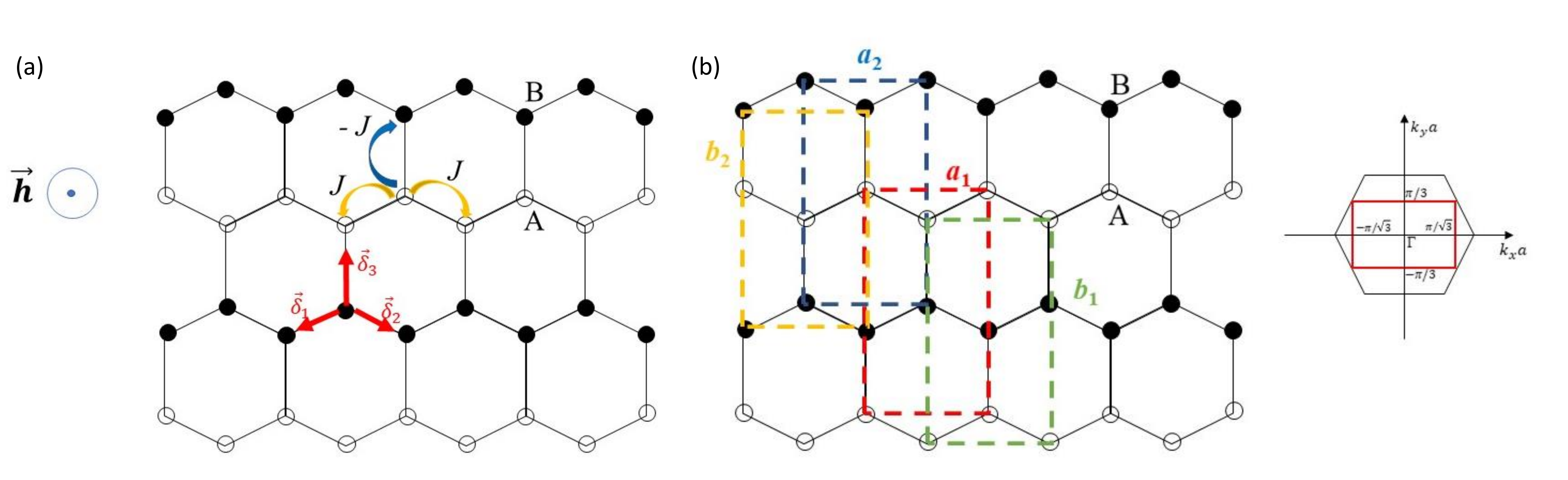}}
\caption{(a) Magnetic structure of a single layer of Fe$^{2+}$ ions in FePS$_3$. The black filled circles represent spins pointing out of the plane of the paper, while the white hollow circles represent spins pointing into the plane of the paper. Two of the nearest neighbors are ferromagnetically coupled, whereas the third is coupled antiferromagnetically. The two crystallographically inequivalent sites are labelled as A and B. For a A-type site, the nearest neighbor bonds are given by the vectors $\Vec{\delta}_1=a\left(-\frac{\sqrt{3}}{2}\hat{x}-\frac{1}{2}\hat{y}\right)$, $\Vec{\delta}_2=a\left(\frac{\sqrt{3}}{2}\hat{x}-\frac{1}{2}\hat{y}\right)$, and $\Vec{\delta}_3=a\hat{y}$, where $a$ is the lattice constant. The magnetic field $\Vec{h}$ points out of the plane of the paper. (b) The structure can be interpreted to made up of 4 interpenetrating magnetic sublattices. Inset: The first magnetic Brillouin zone (in red) and the first crystallographic Brillouin zone (in black).
\label{fig1}}
\end{figure}

\begin{figure}[h!]
\centerline{\includegraphics[scale=0.45, clip]{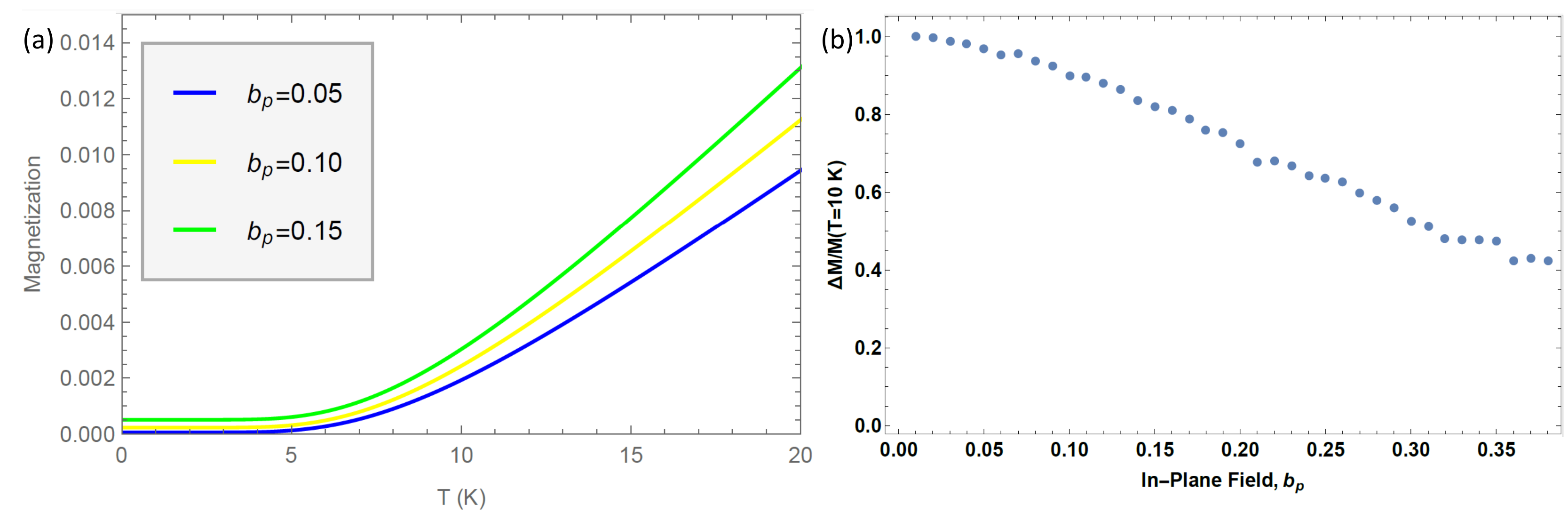}}
\caption{(a) Variation of the total magnetization with temperature for three different in-plane magnetic field strengths. At very low temperatures, the remnant magnetization increases with increase in the strength of the in-plane field. In (b), the relative change in magnetization is seen to decrease with increasing in-plane field strength, $b_p$.
\label{fig2}}
\end{figure}

Here, we explore the effect of magnetic field induced quantum fluctuation at low temperature as an origin of dielectric anomaly observed in the experiment. With the aim of doing so, in addition to the external magnetic field in the out-of-plane direction (which is taken to be the $z$-axis), $B_z=\mu_B g h$, we consider a small in-plane field component along the $y$-axis, $B_p$. In such case, the effective Hamiltonian of the system reads as:
\begin{equation}
    H=-J\sum_{i\in A}\left( \Vec{S}_i.\Vec{S}_{i+\delta_1} + \Vec{S}_i.\Vec{S}_{i+\delta_2} - \Vec{S}_i.\Vec{S}_{i+\delta_3} \right ) - \Delta \sum_i (S_i^z)^2 - B_z \sum_i S_i^z - B_p \sum_i S_i^y
    \label{eqn:hamiltonian}
\end{equation}
where $J>0$ and $\Delta>0$ denote the nearest neighbor exchange interactions and the z-axis anisotropy respectively as discussed earlier. It can be shown that in the absence of the in-plane field, the classical ground state of this model hamiltonian corresponds to a FePS$_3$ kind of order as long as $B_z<JS$ and $\Delta$ is above a small threshold value. Fig.  \ref{fig1}(a) shows a schematic structure of a single layer of the compound, outlining the interactions therein.\\

The classical ground state of this system can be obtained by minimizing the energy with respect to the tilt angles, $\theta_1$ and $\theta_2$ of the up and down spins with respect to the positive and the negative $z$-axis respectively. The configuration so obtained for a small in-plane field component ($B_p \ll B_z$) is 
\begin{eqnarray}
\theta_1 = \frac{b_p (b_z-2\delta)}{1+b_z^2 - (1+2\delta)^2} \nonumber\\
\theta_2 = -\frac{b_p (b_z+2\delta)}{1+b_z^2 - (1+2\delta)^2}
\label{eqn:classicalangles}
\end{eqnarray}
where $b_p=\frac{B_p}{JS}$, $b_z=\frac{B_z}{JS}$ and $\delta=\frac{\Delta}{J}$ refer to the scaled parameters.\\

In order to study the low-lying magnon excitations, we first perform a rotation of axis so that the local $z$-axis for each spin is aligned along its classical direction, as obtained from equation (\ref{eqn:classicalangles}). For the up (down) spins, the transformed spin variables read
\begin{eqnarray}
\Tilde{S}_{z(y)} = S_{z(y)} \cos{\theta_{1(2)}} + S_{y(z)}\sin{\theta_{1(2)}}\nonumber\\
\Tilde{S}_{y(z)} = -S_{z(y)} \sin{\theta_{1(2)}} + S_{y(z)}\cos{\theta_{1(2)}}\nonumber\\
\Tilde{S}_x = S_x
\label{eqn:spinrot}
\end{eqnarray}
We consider the non-Bravais lattice to be composed of four interpenetrating magnetic sublattices as shown in Fig. \ref{fig1}(b) and carry out the following Holstein-Primakoff transformations with respect to the transformed variables\\

\begin{minipage}{.5\textwidth}
\textbf{Sublattice $a_1(b_1)$}:
 \begin{eqnarray*}
     \Tilde{S}_{j,A(B)}^{z}=-S+b_{j,A(B)}^{\dagger} b_{j,A(B)}\\
     \Tilde{S}_{j,A(B)}^{x}=\sqrt{\frac{S}{2}}\left (b_{j,A(B)}+b_{j,A(B)}^{\dagger}\right )\\
     \Tilde{S}_{j,A(B)}^{y}=\frac{1}{i}\sqrt{\frac{S}{2}}\left( b_{j,A(B)}^{\dagger}-b_{j,A(B)}\right )
 \end{eqnarray*}
\end{minipage}
\begin{minipage}{.5\textwidth}
\textbf{Sublattice $a_2(b_2)$}:
 \begin{eqnarray*}
     \Tilde{S}_{j,A(B)}^{z}=S-a_{j,A(B)}^{\dagger} a_{j,A(B)}\\
     \Tilde{S}_{j,A(B)}^{x}=\sqrt{\frac{S}{2}}\left (a_{j,A(B)}+a_{j,A(B)}^{\dagger}\right )\\
     \Tilde{S}_{j,A(B)}^{y}=\frac{1}{i}\sqrt{\frac{S}{2}}\left (a_{j,A(B)}-a_{j,A(B)}^{\dagger}\right )
 \end{eqnarray*}
\end{minipage}\\

where $a_j$ and $b_j$'s obey bosonic commutation relations.\\

The Hamiltonian can be diagonalized using standard Bogoliubov transformation to express the low-lying excitations above the ground state in terms of the magnon variables. For a low enough temperature $T$, a Bose-Einstein distribution can be associated with the concentration of each of the magnon quasiparticles viz
\begin{equation}
    \langle \alpha_k^{\dagger} \alpha_k \rangle = \frac{1}{\exp(\beta \epsilon_k)-1}
    \label{eqn:boseeinstein}
\end{equation}

The temperature dependence of the magnetization $M_z=\sum_j S_j^z$ can be calculated by expressing the spin variables in terms of the magnons and inserting this temperature dependence of the magnon concentration. We plot this variation of the magnetization for three different strengths of the in-plane field component in Fig. \ref{fig2}(a) and in Fig. \ref{fig2}(b), we show how the relative change in the magnetization varies with the strength of the in-plane magnetic field. For the purpose of Fig. \ref{fig2}(b), we define the relative change in magnetization as $\frac{M(T=10\text{ } K)-M(T=0\text{ } K)}{M(T=10\text{ } K)}$.